\documentclass[aps,floatfix,prl,superscriptaddress,reprint,showpacs,10pt,preprintnumbers,longbibliography]{revtex4-2}
\usepackage[utf8]{inputenc}
\usepackage[pdftex]{graphicx}
\usepackage{float}
\usepackage{amsmath}
\usepackage{amssymb}
\usepackage{amsfonts}
\usepackage{dsfont}
\usepackage{array}
\usepackage{bm}
\usepackage{mathrsfs}
\usepackage{pifont}
\usepackage{multirow}
\usepackage{upgreek}
\usepackage{xcolor}
\usepackage{standalone}
\usepackage[pdftex,
            pdftitle={Ruling out the massless up-quark solution to the strong CP problem by computing the topological mass contribution with lattice QCD},
            pdfauthor={Constantia Alexandrou, Jacob Finkenrath, Lena Funcke, Karl Jansen, Bartosz Kostrzewa, Ferenc Pittler, Carsten Urbach},
            bookmarks,
            colorlinks,
            linkcolor=myblue,
            citecolor=mymagenta,
            menucolor=black,
            urlcolor=myblue,
            plainpages=false,
            pdfpagelabels,
            hypertexnames=false]{hyperref}
\usepackage{verbatim}
\usepackage{slashed}
\usepackage{hyperref}
\definecolor{mymagenta}{RGB}{200, 0, 100}
\definecolor{myblue}{RGB}{45, 48, 146}

\begin{document}
\title{Ruling Out the Massless Up-Quark Solution to the Strong CP Problem\\ by Computing the Topological Mass Contribution with Lattice QCD}
\author{Constantia Alexandrou}
\affiliation{Department of Physics, University of Cyprus, P.O. Box 20537, 1678 Nicosia, Cyprus}
\affiliation{Computation-Based Science and Technology Research Center, The Cyprus Institute, 20 Konstantinou Kavafi Street, 2121 Nicosia, Cyprus}
\author{Jacob Finkenrath}
\affiliation{Computation-Based Science and Technology Research Center, The Cyprus Institute, 20 Konstantinou Kavafi Str., 2121 Nicosia, Cyprus}
\author{Lena Funcke}
\affiliation{Perimeter Institute for Theoretical Physics, 31 Caroline Street North, Waterloo, ON N2L 2Y5, Canada}
\author{Karl Jansen}
\affiliation{NIC, DESY Zeuthen, Platanenallee 6, 15738 Zeuthen, Germany}
\author{Bartosz Kostrzewa}
\affiliation{Helmholtz Institut f\"ur Strahlen- und Kernphysik, University of Bonn, Nussallee 14-16, 53115 Bonn, Germany}
\affiliation{Bethe Center for Theoretical Physics, University of Bonn, Nussallee 12, 53115 Bonn, Germany}
\author{Ferenc Pittler}
\affiliation{Helmholtz Institut f\"ur Strahlen- und Kernphysik, University of Bonn, Nussallee 14-16, 53115 Bonn, Germany}
\affiliation{Bethe Center for Theoretical Physics, University of Bonn, Nussallee 12, 53115 Bonn, Germany}
\author{Carsten Urbach}
\affiliation{Helmholtz Institut f\"ur Strahlen- und Kernphysik, University of Bonn, Nussallee 14-16, 53115 Bonn, Germany}
\affiliation{Bethe Center for Theoretical Physics, University of Bonn, Nussallee 12, 53115 Bonn, Germany}

\date{\today}
\begin{abstract}
  The infamous strong CP problem in particle physics can in principle
  be solved by a massless up quark. In particular, it was hypothesized
  that topological effects could substantially contribute to the
  observed nonzero up-quark mass without reintroducing CP
  violation. Alternatively to previous work using fits to chiral perturbation theory, in this Letter, we bound
  the strength of the topological mass contribution with direct 
  lattice QCD simulations, by computing the dependence of the pion mass on the dynamical strange-quark mass. We find that the size
  of the topological mass contribution is inconsistent with the
  massless up-quark solution to the strong CP problem.
\end{abstract}

\maketitle

\textit{Introduction.}---One of the unsolved puzzles in particle physics
is the so-called strong CP problem, where CP stands for the combined
charge conjugation and parity symmetry. In quantum chromodynamics
(QCD), which is the theory of strong interactions, the nontrivial topological vacuum structure generates a CP-violating term
\[
\propto \theta\, G_{\mu\nu}\tilde{G}^{\mu\nu}
\]
in the Lagrangian,
where $\theta$ is an \textit{a priori} unknown
parameter, $G$ is the gluon field strength tensor ,and
$\tilde{G}$ is its dual. However,
experimentally, there is no sign of CP violation in QCD. Instead, the
strong upper bound $\theta\lesssim 10^{-10}$~\cite{Crewther:1979,Afach:2015sja,Abel:2020} from
measurements of the neutron electric dipole moment leads to a severe
fine-tuning problem. 

There are several proposals to overcome this problem, for instance, by
postulating the existence of an axion~\cite{Peccei1977,Weinberg1977,Wilczek1978}. A simple alternative could be
the vanishing of the up-quark mass $m_u$, which at first sight seems
inconsistent with results of current algebra. 
However, Refs.~\cite{Georgi1981,Kaplan1986,Choi1988,Banks1994} pointed out that the
up-quark mass
 in the chiral Lagrangian 
has two different
contributions: a CP-violating perturbative contribution $m_u$ and a
CP-conserving nonperturbative contribution $m_{\rm eff}$ from
topological effects, such as instantons. While $m_u=0$ could be easily
ensured by an accidental symmetry~\cite{Leurer1992,Leurer1993,
  Banks1994,Nelson1996,Kaplan1998}, $m_{\rm eff}$ does not contribute
to the neutron electric dipole moment and is parametrically of order
$m_{\rm eff}\sim m_dm_s/\Lambda_{\rm QCD}$, plausibly as large as the
total required up-quark mass. Testing this simple solution to the
strong CP problem is particularly important because the other proposed
solutions, including the
axion~\citep{Peccei1977,Weinberg1977,Wilczek1978} and
Nelson-Barr~\cite{Nelson1983,Barr1984} mechanisms, face several
theoretical challenges~\cite{Dine:2015jga}. 

As the only tool to reliably test the $m_u=0$
proposal~\cite{Banks1994}, lattice gauge theory has determined the
up-quark mass to $m_u(2~{\rm GeV})\sim 2$~MeV by fitting the light meson spectrum with errors around 5\%
(see Refs.~\cite{Bazavov:2017lyh,Bazavov:2018omf} and Ref.~\cite{Aoki:2019cca} for a review). As proposed in Refs.~\cite{Cohen1999,Dine:2014dga}, it would be beneficial to perform a complementary analysis by calculating the dependence of the pion mass on the dynamical strange-quark mass while keeping the light quark masses fixed. This direct calculation  would have the advantage of avoiding any fitting procedures.

Lattice QCD simulations  
are now being performed, taking into account the first two quark generations
as dynamical degrees of freedom. 
In addition, simulations are performed at (or very close to) 
the physical values of the pion, kaon, and $D$-meson masses \footnote{Note that, in this work, we can use larger than physical quark masses due to reasons discussed below Eq.~\eqref{eq:beta}.}; and  
at various values of the lattice spacing and volumes, such that
systematic effects can be studied and eventually
controlled~\cite{Bazavov:2012xda,Alexandrou:2018egz}.  
Finally, the theoretically sound definitions of the topological 
charge and susceptibility on the lattice (see Ref.~\cite{Alexandrou:2017hqw} for a review) 
allow for directly accessing topological effects related to $m_{\rm eff}$.

In this Letter, we perform a cross-check of the  $m_u>0$ hypothesis based
on the proposals of Refs.~\cite{Cohen1999,Dine:2014dga}. In
particular, we compute the parameter $\beta_2/\beta_1$, which measures
the strength of $m_{\rm eff}$ and probes the contribution of small
instantons and other topological effects to the chiral
Lagrangian. While $\beta_2/\beta_1$ is usually obtained from a
combination of low-energy constants~\cite{Nelson2003}, this indirect lattice method requires
chiral perturbation theory ($\chi$PT). Using \textit{direct} lattice computations instead, we obtain 
the result
$\beta_2/\beta_1 = 0.63(39)~{\rm GeV}^{-1}$
by computing the dependence of the pion mass on the strange-quark
mass. 
Since a bound significantly smaller than $5~{\rm GeV}^{-1}$ provides an exclusion
of the massless up-quark hypothesis~\cite{Banks1994,Dine:2014dga}, our
result rules out this hypothesis, in accordance with previous fits of $\chi$PT to lattice data~\cite{Aoki:2019cca,Nelson2003,Cline1989,Dragos:2019oxn}.

\textit{Method.}---We test
the $m_u=0$ proposal by investigating the variation of the pion mass 
with respect to the strange-quark mass. The general form of the quark-mass dependence of the pion mass reads \footnote{Note that $\beta_1$ is dimensionless as in Ref.~\cite{Banks1994}, while the dimensionful $\beta_2$ translates to $\beta_2 / \Lambda_{\rm \chi SB}$ in Ref.~\cite{Banks1994}.}
\begin{equation}
M_\pi^2 = \beta_1 (m_u+m_d) + \beta_2 m_s (m_u+m_d)+\textnormal{higher orders}\,,
\label{eq:mpisq}
\end{equation}
where the first term is the first-order contribution of the light quark masses in $\chi$PT. The second term receives contributions both from
small instantons that could mimic a nonzero $m_u$ and from higher-order terms in $\chi$PT that are proportional to $m_s$, including logarithmic corrections. In order 
to let topological effects explain the observed value for $m_u$ and to
allow for a solution of the strong CP problem, $\beta_2/\beta_1\approx
5~\mathrm{GeV}^{-1}$ at renormalization scale $\bar\mu =
2~\mathrm{GeV}$ in the $\overline{\textrm{MS}}$ scheme is
required~\cite{Dine:2014dga}.

The most precise and computationally challenging test of the ratio
$\beta_2/\beta_1$ is to vary either the strange-quark mass or the light quark mass, $m_u=m_d\equiv m_\ell$.
For example, by varying $m_s$ while keeping $m_\ell$ 
fixed, we obtain~\cite{Cohen1999,Dine:2014dga}
\begin{equation}
 \frac{\beta_2}{\beta_1}\ \approx\ \frac{M_{\pi,1}^2 - M_{\pi,2}^2}{m_{s,1}
   M^2_{\pi,2} -m_{s,2} M^2_{\pi,1}}\, , 
 \label{eq:beta}
\end{equation}
where $M_{\pi,i}=M_\pi(m_{s,i})$ is the average pion mass as a function of
the varied strange-quark mass $m_{s,i}$ at fixed $m_\ell$. Note that the approximate
result for $\beta_2/\beta_1$ in Eq.~\eqref{eq:beta} is independent of
the up and down quark masses. Crucially, this allows us to reliably compute $\beta_2/\beta_1$ even at larger than physical quark masses. The higher-order
corrections in Eq.~\eqref{eq:mpisq} reintroduce a small residual pion-mass
dependence for $\beta_2/\beta_1$ that finally needs to be cancelled by a chiral
extrapolation. 

While this challenging
direct method to compute the ratio $\beta_2/\beta_1$ is independent of $\chi$PT, the more common indirect
method is to use the chiral
Lagrangian. For example, Ref.~\cite{Dine:2014dga} used lattice data
from the Flavour Lattice Averaging Group (FLAG) report of 2013~\cite{Aoki2013} to estimate
$\beta_2/\beta_1 \simeq (1\pm 1)~{\rm GeV}^{-1}$ neglecting chiral logarithms
and higher-order terms in the chiral Lagrangian. To check the
consistency of our computations with the results of
Ref.~\cite{Dine:2014dga}, we have also computed $\beta_2 /\beta_1$
indirectly by using chiral fits and measuring
$M_K^2(m_s)$, obtaining excellent agreement with Ref.~\cite{Dine:2014dga}.

\begin{table}
  \centering
  \begin{tabular*}{\linewidth}{@{\extracolsep{\fill}}lrr}
    \hline\hline \\[-2.0ex]
    Ensemble & $M_\pi~[\mathrm{MeV}]$ & 
    $m_s~[\mathrm{MeV}]$\\
    \hline\hline  \\[-2.0ex]
    A60 &  $386(16)$ & $98(4)$ \\
    A60s&  $387(16)$ & $79(4)$ \\
    A80 &  $444(18)$ & $98(4)$\\
    A80s&  $443(18)$ & $79(4)$ \\
    A100 & $494(20)$ & $100(4)$ \\ 
    A100s& $495(20)$ & $79(4)$ \\ 
    \hline  \\[-2.0ex]
    cA211.30.32  & $276(3)$ & $99(2)$\\
    cA211.30.32l & $275(3)$ & $\phantom{0}94(2)$\\
    cA211.30.32h & $276(3)$ & $104(2)$\\
    \hline\hline
  \end{tabular*}
  \caption{Pion- and strange-quark masses in physical units for the
    ensembles used in this work. The strange-quark mass is quoted at
    $2~\mathrm{GeV}$ in the $\overline{\mathrm{MS}}$ scheme.}
  \label{tab:cA211}
\end{table}

\textit{Lattice computation.}---In this Letter, we use gauge
configurations generated by the Extended Twisted Mass Collaboration
(ETMC) with the Iwasaki gauge action~\cite{Iwasaki:1985we} and Wilson
twisted mass fermions at maximal
twist~\cite{Frezzotti:2000nk,Frezzotti:2003xj} with up, down, strange, 
and charm dynamical quark flavors. Up and down quarks are mass
degenerate. 
All the gauge configuration ensembles we used are listed together with
the corresponding pion- and strange-quark mass values in Table~\ref{tab:cA211}. 
For details on how these values are
obtained, we refer to the Supplemental Material~\cite{SupplMat}.

We first perform the analysis using three sets each with a pair of ensembles (AX and AXs with $X=60, 80,100$) without the
so-called clover term in the action. Details on the production of these ensembles can be found in
Ref.~\cite{Baron:2010bv}. Each pair with $X=60$, $80$, and $100$ has 
identical parameters apart from strange- and 
charm-quark-mass values, which are close to their physical
values. The three pairs have equal strange- and charm-quark masses
within errors but differ in the light quark-mass value corresponding
to unphysically large pion-mass values of 
about $386~\mathrm{MeV}$, $444~\mathrm{MeV}$ and
$494~\mathrm{MeV}$, respectively. The lattice spacing value
corresponds to $a = 0.0885(36)\ \mathrm{fm}$~\cite{Carrasco:2014cwa}
determined from the pion decay constant $f_\pi$. 

In addition, we use one ensemble (cA211.30.32) that includes the
clover term in the action~\cite{Alexandrou:2018egz}. While Wilson twisted mass fermions
at maximal twist automatically remove discretization effects linear in
the lattice spacing $a$~\cite{Frezzotti:2003ni} and thus leave only 
lattice artifacts at $\mathcal{O}(a^2)$, the clover term reduces these $\mathcal{O}(a^2)$ effects even further~\cite{Abdel-Rehim:2015pwa}. The cA211.30.32 ensemble
has a smaller pion-mass value of about $270~\mathrm{MeV}$ as well as
strange- and charm-quark mass values again close to their physical
values. The lattice spacing value is $a=0.0896(10)\ \mathrm{fm}$ determined using the nucleon mass dependence on the pion mass.
This estimation is done by employing $\chi$PT at 
$\mathcal{O}(p^3)$ \cite{Gasser:1987rb,Tiburzi:2008bk}, where $p$ is a typical meson momentum.
Similarly to Ref.~\cite{Alexandrou:2018egz}, the $N_f=2+1+1$ nonclover twisted mass ensembles \cite{Alexandrou:2014sha} 
at different lattice spacings, which also include the AX ensembles, were used to control the chiral extrapolations.

Since the pion mass of the cA211.30.32 ensemble is significantly smaller than the ones of the AX
 ensembles and thus closer 
 to the physical value, we consider this ensemble as the most appropriate one  to compute our final value of $\beta_2/\beta_1$. Moreover, cA211.30.32 uses the same  action
 as ensembles that are currently under production with a physical value of the  pion mass. These ensembles  
could be used, in principle, in future work to repeat the
calculation presented here at the physical point.
For cA211.30.32, we have simulations for only one dynamical strange-quark mass; thus, it is necessary to apply the so-called {\em
  reweighting technique} to investigate the strange-quark-mass
dependence of $M_\pi$ while keeping the charm and light quark masses constant~\cite{SupplMat}.
We denote with cA211.30.32l (cA211.30.32h) the reweighted ensemble with a 5\% lower (higher) strange-quark-mass value than the original ensemble cA211.30.32.

In contrast, for the AX(s) ensembles, we have pairs of ensembles with different
dynamical strange-quark masses; thus, we can use a \textit{direct
  approach} to investigate the strange-quark-mass
dependence of $M_\pi$. Note that in this case also, the charm quark mass differs
slightly, but its value is so close to the cutoff that this difference will not affect our results. While the AX(s)
ensembles have rather heavy pion masses (see Table~\ref{tab:cA211}), they are ideal to test the robustness of the reweighting 
procedure that we apply to cA211.30.32. In fact, we use these ensembles to demonstrate that reweighting
works successfully~\cite{SupplMat}.
In addition, the $\beta_2/\beta_1$ values from these ensembles provide an insight into the 
pion-mass dependence of $\beta_2/\beta_1$.

\begin{table}
  \centering
  \begin{tabular*}{.49\textwidth}{@{\extracolsep{\fill}}lccc}
    \hline\hline  \\[-2.0ex]
    Ensemble & $\beta_2~[\mathrm{GeV^2}]$ &
    $\beta_1~[\mathrm{GeV^3}]$ &
    $\beta_2/\beta_1~[\mathrm{GeV}^{-1}]$\\
    \hline\hline  \\[-2.0ex]
    A60(s)  & $-0.0009(08)$ & $\phantom{-}0.0029(4)$ & $-0.32(26)$\\
    A80(s)  & $\phantom{-}0.0005(10)$ & $\phantom{-}0.0036(4)$ & $\phantom{-}0.15(30)$\\
    A100(s) & $-0.0010(10)$ & $\phantom{-}0.0053(6)$ & $-0.19(19)$\\
    \hline \\[-2.0ex]
    cA211.30.32(h)   & $0.00007(11)$ & $0.00039(5)$ & $\phantom{-}0.18(30)$ \\
    cA211.30.32(l)   & $0.00026(11)$ & $0.00037(5)$ & $\phantom{-}0.69(33)$ \\
    cA211.30.32(h,l) & $0.00033(12)$ & $0.00076(5)$ & $\phantom{-}0.43(16)$ \\
    \hline\hline
  \end{tabular*}
  \caption{Results for $\beta_2$, $\beta_1$ and $\beta_2/\beta_1$ from Eq.~(\ref{eq:beta}) in physical units for all ensembles at
    $\bar\mu=2~\mathrm{GeV}$ in the 
    $\overline{\mathrm{MS}}$ scheme.}
  \label{tab:beta12}
\end{table}

\textit{Results.}---Using the values of $M_\pi$ and $m_s$
from Table~\ref{tab:cA211} as input (more precisely, the corresponding values in lattice units~\cite{SupplMat}), we compute $\beta_2/\beta_1$ from 
Eq.~(\ref{eq:beta}). The results from this direct approach for the
three pairs A60(s), A80(s) and A100(s) are compiled in 
Table~\ref{tab:beta12}, where we quote $\beta_2/\beta_1$ as well as $\beta_1$ and $\beta_2$ separately.
Since the pion-mass differences are all zero within errors (see
Table~\ref{tab:cA211}), we find that
$\beta_2/\beta_1$ is compatible with zero as well. Note that the errors of
the observables compiled in Table~\ref{tab:cA211} are correlated per
ensemble. This correlation is taken into account in our analysis for
$\beta_2/\beta_1$. 

Finally, we use reweighting on the cA211.30.32 ensemble to vary the
strange-quark mass by $\pm5\%$ around its original value. 
The change in the pion mass with the strange-quark mass
is not significant; see Table~\ref{tab:cA211}. The corresponding
values for $\beta_2/\beta_1$~\cite{SupplMat} are again compiled in
Table~\ref{tab:beta12}. Here, we denote with cA211.30.32(h) the
value for $\beta_2/\beta_1$ obtained from the combination of the ensembles
cA211.30.32h and cA211.30.32. Likewise, cA211.30.32(l) is the
combination of cA211.30.32 and cA211.30.32l, while cA211.30.32(h,l) is the combination of cA211.30.32h and cA211.30.32l.

\begin{figure}[t]
  \centering
  \includegraphics[width=0.48\textwidth]{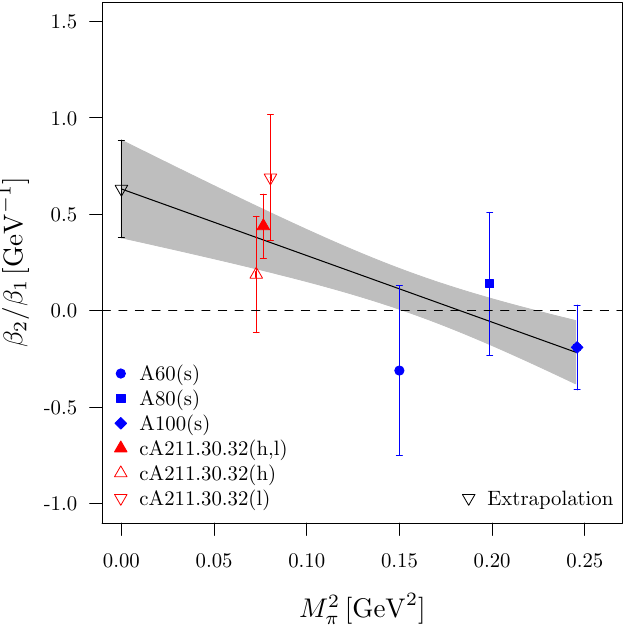} 
  \caption{The ratio $\beta_2/\beta_1$ as a function of the squared
    pion mass $M_\pi^2$ in physical units. The solid line with the
    $1\sigma$ error band represents a linear extrapolation in
    $M_\pi^2$. We extrapolate to the chiral limit to eliminate higher-order corrections to $\beta_2/\beta_1$; see Eqs.~\eqref{eq:mpisq} and \eqref{eq:beta}.}
  \label{fig:beta21}
\end{figure}

In Fig.~\ref{fig:beta21}, we show the values of the ratio
$\beta_2/\beta_1$ at $\bar\mu=2~\mathrm{GeV}$ in the
$\overline{\textrm{MS}}$ scheme as a function of the squared pion mass 
$M_\pi^2$ in physical units. The
three blue points at heavier pion-mass values correspond to the three
pairs of the AX(s) ensembles without the clover term. The three red points at
lower pion-mass values correspond to the cA211.30.32 ensemble including the clover
term. The latter three points are slightly displaced horizontally for
better legibility. While all of the points are compatible with zero at
the $1.5\sigma$ level, we observe a slight trend toward larger
$\beta_2/\beta_1$ values with decreasing pion-mass values. 

In addition, we show in Fig.~\ref{fig:beta21} a linear extrapolation
of $\beta_2/\beta_1$ in $M_\pi^2$ to the chiral limit. This linear
dependence can be justified with $\chi$PT, which
predicts~\cite{Aoki:2019cca} 
\begin{equation}
  \frac{\beta_2}{\beta_1} \approx \frac{\alpha_2}{\alpha_1 +(\alpha_3
    /\alpha_1)M_\pi^2} \approx \frac{\alpha_2}{\alpha_1} -
  \frac{\alpha_2\alpha_3}{\alpha_1^3}M_\pi^2 \label{chPT}
\end{equation}
modulo logarithmic corrections, where $\alpha_{1,2,3}$ are combinations of low-energy constants with
$\alpha_1\gg (\alpha_3/\alpha_1)M_\pi^2$, and $M_\pi^2=\alpha_1 m_\ell
+\mathcal{O}(\alpha_{2,3})$ with $\mathcal{O}(\alpha_{2,3})/ (\alpha_1
m_\ell)\approx 0.1$. Since the data points for the ensemble cA211.30.32
are highly correlated, we include only the combination of cA211.30.32h
and cA211.30.32l in the fit denoted as cA211.30.32(h,l) in
Table~\ref{tab:beta12}. The fit has $\chi^2/\mathrm{dof}=3.28/2$ (i.e., a $p$-value of $0.2$) and the chirally extrapolated value reads
$\beta_2/\beta_1 = 0.63(25)~\mathrm{GeV}^{-1}$. As mentioned, we
extrapolate to the chiral limit to cancel the residual pion-mass
dependence in Eq.~\eqref{chPT}, which stems from higher-order
corrections in Eq.~\eqref{eq:mpisq} and does
not appear in the expression for $\beta_2/\beta_1$ in
Eq.~\eqref{eq:beta}. Our data thus confirm in hindsight that the
approximation in Eq.~\eqref{eq:beta} is justified. 

\textit{Discussion.}---All the estimates for the ratio
$\beta_2/\beta_1$ presented in this Letter are consistent with
zero at the $1.5\sigma$ level. With the chiral extrapolation explained
above and $1\sigma$ statistical uncertainty, we exclude a
value of $5~\mathrm{GeV}^{-1}$ by an amount significantly larger than $10\sigma$. The remaining question is 
whether there are additional systematic uncertainties that could
potentially spoil this conclusion. 

Let us first consider the discretization
errors for
$\beta_2/\beta_1$, which are of order $(a\Lambda_\mathrm{QCD})^2$ multiplied by an unknown coefficient, with
$\Lambda_\mathrm{QCD}=341(12)~\mathrm{MeV}$~\cite{Bruno:2017gxd}. We can reliably estimate the coefficient by using the known continuum
extrapolation values for $M_\pi^2$ and $m_s$ for the AX ensembles~\cite{Carrasco:2014cwa}. By comparing these continuum values to our lattice results for $M_\pi^2$ and $m_s$, we can infer the size of
discretization errors at our given lattice spacing. Depending
on the scaling variable, the discretization errors in $M_\pi^2$ and
the strange-quark mass are both on the order of $5-10\%$.
If propagated generously, this implies a $10\%$
uncertainty on the numerator, a $15\%$ uncertainty on the
denominator, and thus about $20\%$ on the ratio
$\beta_2/\beta_1$. Note that this estimate is highly conservative because most of the discretization effects cancel in the 
differences in both the numerator and the denominator. Because of the
reduced lattice artifacts with the action including the clover term
(see Supplementary Material~\cite{SupplMat}), we do not expect larger uncertainties on
the ratio for the ensemble cA211.30.32 stemming from discretization
effects. 

In addition, there is a residual pion-mass dependence of $\beta_2/\beta_1$, which we account for 
by extrapolating to the chiral limit. In this extrapolation, the
errors stemming from different lattice artifacts of the AX and
cA211.30.32 ensembles are
taken into account by the above-mentioned $20$\% uncertainty. Last, there are finite-size effects for $M_\pi$ proportional
to $\exp(-M_\pi L)$, with $L$ as the spatial extent of the lattice, but
no finite-size corrections to $m_s$. Since the strange-quark-mass dependence of $M_\pi$ is so weak, these finite-size effects are equal for $M_{\pi,1}^2$ and $M_{\pi,2}^2$, and thus
cancel in the ratio $\beta_2/\beta_1$.

In summary, taking the chirally extrapolated value for
$\beta_2/\beta_1$ plus the $1\sigma$ statistical error and the
$20\%$ uncertainty for discretization
effects, we arrive at the following conservative estimate: 
\begin{equation}
  \begin{split}
   \frac{\beta_2}{\beta_1}\ &=\ 0.63(25)_\mathrm{stat}(14)_\mathrm{sys}~\mathrm{GeV}^{-1}\\
    &=\ 0.63(39)~\mathrm{GeV}^{-1}\\
  \end{split}
  \label{eq:bound}
\end{equation}
at $\bar\mu=2~\mathrm{GeV}$ in the $\overline{\textrm{MS}}$
scheme. For the final estimate we have added the errors linearly.
Note that our data are equally well compatible with a
constant extrapolation in $M_\pi^2$, which would lead to a significantly
smaller value at the physical point. Thus, we consider 
Eq.~(\ref{eq:bound}) as a conservative estimate. Moreover, the logarithmic corrections from chiral perturbation theory contributing to $\beta_2/\beta_1$~(see, e.g., Refs.~\cite{Novikov1981,Gasser1985}) are of the same order as our value in Eq.~\eqref{eq:bound}; therefore, the topological contribution to $\beta_2/\beta_1$ should be even smaller.

\textit{Conclusion.}---In this Letter, we have tested the massless up-quark
solution to the strong CP problem by directly investigating the
strange-quark-mass dependence of the pion mass on the lattice. This allows us to
determine the ratio $\beta_2/\beta_1$, which would need to be larger
than $5~\mathrm{GeV}^{-1}$ to solve the strong CP problem.

Since all our estimates of $\beta_2/\beta_1$ are compatible with zero,
we obtain a strong upper bound for $\beta_2/\beta_1$
including residual uncertainties stemming from discretization errors
and chiral extrapolation. The result in Eq.~(\ref{eq:bound}) is clearly
incompatible with the massless up-quark solution to the strong CP
problem. This exclusion of the $m_u=0$ solution is consistent with previous results using $\chi$PT and direct fits of the light meson spectrum.

Given our conservative error estimates, we consider it highly
unlikely that the factor of 5 needed to rescue the solution to the
strong CP problem is hidden in the quoted uncertainties. A
confirmation of this result  using ensembles with physical pion-mass values could be undertaken in the future, once different values
for the lattice spacing become available for a continuum
extrapolation.

Our direct lattice results also quantitatively support the large-$N$ picture as a
good description of QCD at low scales, because the 
coefficient of the nonperturbatively induced mass operator is known to
be suppressed in the large-$N$ limit
\citep{Banks1994,Cohen1999}. Thus, our computations reliably
demonstrate that the topological vacuum contributions to the chiral
Lagrangian are negligible. \\

\begin{acknowledgments}
  We thank all members of the ETM Collaboration for the most enjoyable
  collaboration. We also thank J.~Gasser, D.~Kaplan, T.~Banks, Y.~Nir, and N.~Seiberg for helpful comments on the draft; and U.-G.~Meißner for useful comments and discussions.
  We kindly thank F.~Manigrasso and K.~Hadjiyiannakou for providing 
  the necessary correlators for computing the nucleon mass on the ensemble
  cA211.30.32. 
  The authors gratefully acknowledge the Gauss Centre for Supercomputing (GCS)
  e.V. (www.gauss-centre.eu) for funding this project by providing
  computing time on the GCS supercomputer JUQUEEN~\cite{juqueen} and the
  John von Neumann Institute for Computing (NIC) for computing time
  provided on the supercomputers JURECA~\cite{jureca} and JUWELS at Jülich
  Supercomputing Centre (JSC) under the projects hch02, ecy00, and hbn28.
  The project used resources of the SuperMUC at the Leibniz Supercomputing
  Centre under the Gauss Centre for Supercomputing e.V.
  project pr74yo.
  The cA211.30.32 ensemble was generated on the Marconi-KNL supercomputer
at CINECA within PRACE project Pra13-3304.
  This project was funded in part by the DFG as a project in the
  Sino-German CRC110 (TRR110) and by the PRACE Fifth and Sixth Implementation Phase 
  (PRACE-5IP and PRACE-6IP) program of the European Commission under
  Grant Agreements No.\ 730913 and No.\ 823767. Research at Perimeter Institute is supported in part by the Government of Canada through the Department of Innovation, Science and Industry Canada and by the Province of Ontario through the Ministry of Colleges and Universities.
  The open source software packages tmLQCD~\cite{Jansen:2009xp},
  Lemon~\cite{Deuzeman:2011wz}, DDalphaAMG~\cite{Frommer:2013fsa,Alexandrou:2016izb,Alexandrou:2018wiv},
  QUDA~\cite{Clark:2009wm,Babich:2011np,Clark:2016rdz}, and R~\cite{R:2005} have been used.\looseness=-1
\end{acknowledgments}

\pagebreak

\appendix
\onecolumngrid
\vspace{\columnsep}
\begin{center}
\textbf{\large Supplemental Material: Applying  Mass Reweighting to the Strange-Quark Mass\\*[0.2em] Keeping Maximal Twist}
\end{center}
\vspace{0.6\columnsep}
\twocolumngrid

We discuss here our procedure to determine the strange-quark mass and explain the  reweighting approach used for the cA211.30.32 ensemble to obtain the probability distribution at different values of the strange-quark mass. 
In Table~\ref{tab:cA211x}, we list the bare parameters of the gauge ensembles used in this work. They consist of $N_f=2+1+1$ twisted-mass ensembles without a clover term for the AX(s) ensembles and with a clover term~\cite{Sheikholeslami:1985ij} for the cA211.30.32 ensemble.
The ensembles without a clover term come in pairs having the same light quark mass, determined by the parameter $\mu_\ell$, and different strange-quark masses, related to 
the parameters $\mu_\sigma$ and  $\mu_\delta$. They are denoted as A60 and A60s for the lightest pion mass, A80 and A80s for the intermediate and A100 and A100s for the heaviest pion mass. They
all have the same lattice spacing $a=0.0885(36)$~fm. 
For the ensemble with a clover term, denoted as cA211.30.30, and with the lattice spacing $a=0.0896(10)$,
we use re-weighting in the strange-quark mass of $\pm 5$\%. We denote the resulting lighter  and heavier mass ensembles as cA211.30.30l and cA211.30.32h, respectively.

\section{Determination of the Strange-Quark Mass}

In order to evaluate the ratio $\beta_2/\beta_1$, we need to determine the renormalised strange-quark mass. Two different approaches are used that provide a cross-check.
The first approach uses the following relation of the bare parameters $\mu_\sigma$ and $\mu_\delta$ of Table~\ref{tab:cA211x}
to the renormalized strange- and charm-quark masses:
\begin{equation}
 m_s^a = \frac{1}{Z_P}\mu_\sigma
 - \frac{1}{Z_S} \mu_\delta  \quad \textrm{and} \quad m_c^a
 = \frac{1}{Z_P}  \mu_\sigma + \frac{1}{Z_S} \mu_\delta\,.
 \label{eq:ms_mc}
\end{equation}
Here, $Z_P$ and $Z_S$ are the pseudoscalar and scalar renormalization
functions. For the ensembles without a clover term, they have the following values~\cite{Carrasco:2014cwa}:
\[
Z_P=0.529(07)\quad \textrm{and} \quad Z_S=0.747(12)\,.
\]
For the ensemble cA211.30.32, we use~\footnote{P. Dimopoulos, private communication.}
\[
Z_P=0.482(5)\quad \textrm{and} \quad Z_S=0.623(5)\,.
\]
The renormalization functions are given at $\bar\mu=2\ \mathrm{GeV}$ in the $\overline{\mathrm{MS}}$ scheme. 
Knowing the values of $Z_P$ and $Z_S$, we can directly
compute $m^a_{s,c}$ from the bare parameters $\mu_\sigma$ and
$\mu_\delta$.

An alternative approach is to use the kaon mass $M_K$ computed in two ways: 
i) we compute $M_K$ within the same fermion discretization using the same bare strange-quark mass for the valence as for the one in the sea.
We denote this kaon mass by $M_K^{\rm unitary}$ since the valence- and the sea-quark masses are the same; 
ii) we compute the kaon mass using Osterwalder-Seiler (OS) valence strange
quarks~\cite{Frezzotti:2004wz} with bare strange-quark mass
$\hat{\mu}_s$. 
The value of $\hat{\mu}_s$ is then adjusted such that the resulting kaon mass $M_K^{\rm OS}$ matches  $M_K^{\rm unitary}$.
The desired value of the bare strange-quark mass $\mu_s$ is given by 
\begin{equation}
\mu_s = \hat{\mu}_s\ \bigl|\bigr.\ _{ M_K^\mathrm{unitary}(\mu_\sigma, \mu_\delta) =
M_K^\mathrm{OS}(\hat{\mu}_s)}\,,
\end{equation}
yielding the renormalised strange-quark mass
\begin{equation}
m_s^b\ =\ \frac{1}{Z_P} \mu_s\, ,
\end{equation}
where $Z_P$ is the same pseudoscalar renormalization constant as the one used in the first approach. 

\begin{table}
  \centering
  \begin{tabular}{lcccccc}
    \hline\hline \\[-2.0ex]
    Ensemble & $\beta$ & $L/a$ & $a\mu_\ell$ & $a\mu_\sigma$
    & $a\mu_\delta$ & $aM_\pi$\\
    \hline\hline  \\[-2.0ex]
      A60  & 1.90 & $24$ & $0.006$ & $0.15$ & $0.190$ & 0.17308(32)\\
    A60s &        &      &         &        & $0.197$ & 0.17361(31)\\
    A80  &         &      & $0.008$ &        & $0.190$ & 0.19922(30)\\
    A80s &         &      &         &        & $0.197$ & 0.19895(42)\\
    A100 &         &      & $0.010$ &        & $0.190$ & 0.22161(35)\\
    A100s&         &      &         &        & $0.197$ & 0.22207(27)\\
    \hline  \\[-2.0ex]
    cA211.30.32 & 1.726 & 32 & $0.003$ & $0.1408$ & $0.1521$
    & $0.12530(14)$\\
    cA211.30.32l & & & & $0.1402$ & $0.1529$
    & $0.12509(16)$\\
    cA211.30.32h & & & & $0.1414$ & $0.1513$
    & $0.12537(14)$\\
    \hline\hline
  \end{tabular}
  \caption{Parameters of the ensembles used in this work.  $\beta=6/g_0^2$
is the inverse-squared gauge coupling, $\mu_\ell$ the bare 
light quark mass, and $\mu_\sigma$ and $\mu_\delta$ are parameters related to the renormalized
strange- and charm-quark masses. All dimensionful quantities are given
  in units of the lattice spacing $a$. For the cA211 ensembles, the
  value of the clover improvement coefficient is
  $c_\mathrm{sw}=1.74$.}
  \label{tab:cA211x}
\end{table}

In order to carry out the matching as described above, we compute $M_K^{\rm OS}$  for three
values of  $a\hat{\mu}_s$ and interpolate
$(aM_K^\mathrm{OS})^2$ linearly in $a\hat{\mu}_s$.  The unitary kaon masses are taken from  Ref.~\cite{Ottnad:2017bjt} or computed when not available.
For the ensembles without the clover term, we
use $a\hat{\mu}_s=0.017, 0.019, 0.0225$, and for the cA211.30.32 ensemble,
we use $a\hat{\mu}_s = 0.0176, 0.0220, 0.0264$.

In
Table~\ref{tab:beta21comp}, we give $\beta_2/\beta_1$ computed
using $m_s^a$ and $m_s^b$. As can be seen, the mean values agree very well, albeit with large statistical errors. This indicates that discretization artifacts largely cancel in the ratio.
We thus conclude that either approach can be  utilized to fix the strange-quark mass and proceed with $m_s^b$ to compute the ratio $\beta_2/\beta_1$.

\begin{table}
  \centering
  \begin{tabular*}{\linewidth}{@{\extracolsep{\fill}}lrr}
    \hline\hline \\[-2.0ex]
    Ensemble & $\left(\beta_2/\beta_1\right)^a\ [\mathrm{GeV}^{-1}]$ &
    $\left(\beta_2/\beta_1\right)^b\ [\mathrm{GeV}^{-1}]$ \\
    \hline\hline \\[-2.0ex]
    A60(s)  & $-0.29(24)$ & $-0.32(26)$ \\
    A80(s)  & $+0.13(26)$ & $+0.15(30)$ \\
    A100(s) & $-0.20(19)$ & $-0.19(19)$ \\
 \hline\\[-2.0ex]
    cA211.30.32(h) & $+0.176(291)$ &  $+0.182(300)$  \\
    cA211.30.32(l) & $+0.678(326)$ &  $+0.691(332)$  \\
    cA211.30.32(l,h) & $+0.421(160)$ &  $+0.432(165)$ \\
       \hline\hline
  \end{tabular*}
  \caption{The ratio $\beta_2/\beta_1$ computed from $m_s^a$ and $m_s^b$ in
  inverse GeV at $2$ GeV in the $\overline{\mathrm{MS}}$ scheme.}
  \label{tab:beta21comp}
\end{table}

\section{Reweighting in the Strange-Quark Mass}

For the cA211.30.32 ensemble, we apply mass reweighting~\cite{Hasenfratz:2008fg,Finkenrath:2013soa} to obtain the probability distribution at three different values of the strange-quark mass. 
This works if the mass shift is small compared to the original strange-quark mass $m_s$. Since the two above-mentioned approaches  yield consistent values for $\beta_2/\beta_1$ and the definition of $m_s^a$ in Eq.~(\ref{eq:ms_mc}) is technically easier to use for the re-weighting, we use the $m_s^a$ definition to calculate the resulting shifts in the heavy quark parameters.
To keep the notation tidy, we drop the index $a$ from $m_s^a$ and set the lattice spacing to unity in what follows.

In the Boltzmann weight $W = e^{-S_g} D_f$, a shift in the strange-quark mass only affects the fermionic part
\begin{align}
\begin{split}
 D_f (\tilde{m},\mu_\ell,m_s,m_c)  =\;  &\textrm{det} \left[D_\ell^2(\tilde{m},\mu_\ell)\right]\times\\
 &\textrm{det} \left[D_{\rm ND}(\tilde{m},m_s,m_c)\right] \,,
 \end{split}
\end{align}
where $D_\ell$ and $D_{\rm ND}$ are the light and non-degenerate twisted-mass Wilson Dirac operators, respectively.
The untwisted bare quark mass $\tilde{m}$ is an input parameter which receives an additive renormalization factor $m_{\rm cr}$ and is, when subtracted by $m_{\rm cr}$, proportional to the current quark mass: $m_{\rm PCAC} \propto \tilde{m} - m_{\rm cr}$. The current quark mass can be determined via a suitable ratio of matrix elements via the partially-conserved axial current (PCAC) relation.
It should be noted that $m_{\rm cr}$ implicitly depends on all other bare parameters in the theory.
In the twisted-mass formulation, the condition $m_{\rm PCAC} = 0$ is referred to as \emph{maximal twist} and results in \emph{automatic $\mathcal{O}(a)$-improvement} of all physical observables.

Since we are using the $N_f=2+1+1$ twisted-mass action,
the mass-degenerate light quark part of $D_f$ is $\det D_\ell^2(\tilde{m},\mu_\ell)=D_\ell^\dagger(\tilde{m}) D_\ell(\tilde{m})+\mu_\ell^2$ and depends on $\tilde{m}$, which acts as a constant shift to the diagonal of $D_\ell$, and on the light twisted mass $\mu_\ell$, which acts as a twist in spin space via $\gamma_5$ to the diagonal of $D_\ell$. The non-degenerate heavy quark part is $\det D_{\rm ND}(\tilde{m}, m_s,m_c)$ and depends on $\tilde{m}$ and, of course, the strange- and charm-quark masses, $m_s$ and $m_c$.
The latter are functions of the bare quark mass parameters $\mu_\sigma$ and $\mu_\delta$, as given in Eq.~(\ref{eq:ms_mc}).

Our results presented in the main manuscript are obtained by performing a reweighting in $m_s$ keeping $m_c$ constant. From Eq.~(\ref{eq:ms_mc}), one can rewrite $\mu_\sigma=Z_P m_c-(Z_P/Z_S)\mu_\delta$. This means that a change of the strange-quark mass $m_s$ keeping $m_c$ constant
requires knowledge of the ratio of the pseudoscalar to scalar renormalization functions, $Z_P/Z_S$.
However, a change in $m_s$ can result in a significant change in $m_{\rm cr}$, requiring a corresponding adjustment of $\tilde{m}$ to maintain maximal twist and an absence of discretization artifacts linear in the lattice spacing (see section 3.1 of Ref.~\cite{Alexandrou:2018egz} for more details).
The AX, AXs and cA211.30.32 ensembles have been simulated at maximal twist, while the reweighting procedure to obtain cA211.30.32(l,h) also takes into account the necessary shifts in $\tilde{m}$ as the strange-quark mass is varied.

\begin{figure}
  \centering
  \includegraphics[width=0.48\textwidth]{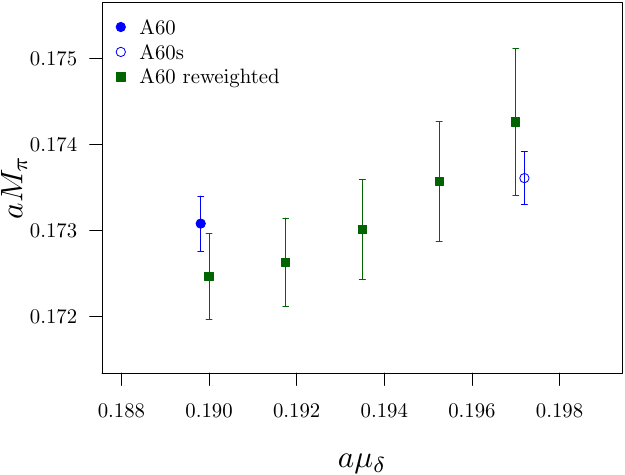} 
  \caption{Pion mass as a function of $a\mu_\delta$ for A60 and
    A60s. The green filled squares correspond to the reweighting analysis of a subset of the gauge configurations of the  A60 ensemble. The
    blue circles, which are slightly displaced horizontally for better legibility, correspond to
    the original measurements on the full A60 (filled blue circle) and A60s
  (open blue circle) ensembles.}
  \label{fig:MpiA60}
\end{figure}

We first confirm that our reweighting for cA211.30.32 works correctly by performing a corresponding reweighting between the parameters of the A60 and A60s ensembles with $M_\pi \sim 400$~MeV.
We vary $\mu_\delta \rightarrow \mu_\delta + \delta \mu_\delta = \mu_\delta'$ keeping $\mu_\sigma$ constant. Note that both $m_s$ and $m_c$ change in this test case, and we account for the change in $m_{\rm cr}$ by suitably reweighting also $\tilde{m} \rightarrow \tilde{m} + \delta \tilde{m} = \tilde{m}'$. 
Reweighting proceeds by correcting the Boltzmann weight with the factor
\begin{equation}
 R_w = R_\ell(\tilde{m},\tilde{m}') R_{\rm ND}(\tilde{m},\tilde{m}',\mu_\delta',\mu_\delta)\,,
 \label{eq:Wdet}
\end{equation}
where
\begin{equation}
 R_{\rm ND}(\tilde{m},\tilde{m}',\mu_\delta,\mu_\delta') = \frac{1}{\textrm{det} \left[D_{\rm ND}(\tilde{m},\mu_\delta) D_{\rm ND}^{-1}(\tilde{m}',\mu_\delta')\right]}
 \label{eq:WND}
\end{equation} 
and
\begin{equation}
  R_\ell(\tilde{m},\tilde{m}') = \frac{1}{\textrm{det} \left[D_\ell^\dagger(\tilde{m}) (D_\ell(\tilde{m}') D_\ell^\dagger(\tilde{m}'))^{-1} D_\ell(\tilde{m})\right]}~.
  \label{eq:Wll}
\end{equation}
We show in Fig.~\ref{fig:MpiA60} the results of the reweighting, starting from the ensemble A60. Since we have performed the reweighting only on a subset of the gauge configurations of A60, we include for comparison the value of the pion mass extracted when using the full ensemble. As expected, the two values are in agreement and the comparison is aimed at having an idea of the error using the subset.  
The determinants for a ratio of matrices $A\in \mathbb{C}^{V\times V}$ with a positive definite partner matrix $A^\dagger + A$ can be computed via a stochastic estimation of the integral representation of the determinant~\cite{Finkenrath:2013soa}
given by
\begin{equation}
 \det A^{-1} = \int \textrm{D}\eta \, e^{-\eta^\dagger A \eta} = \frac{1}{N_{\rm st}}\sum_{i=1}^{N_{\rm st}} \frac{e^{-\chi_i^\dagger A \chi_i}}{e^{-\chi_i^\dagger\chi_i}} + \mathcal{O}(N_{\rm st}^{-1/2})
 \label{eq:A}
\end{equation}
with complex Gaussian distributed fields $\chi_i$, the total number of stochastic estimates $N_{\rm st}$, and with a normalized integral measure $\textrm{D}\eta = \prod_{i \in V} d \textrm{Re}(\eta_i) d \textrm{Im}(\eta_i)/\pi$, where $\eta_i$ is the $i$th entry of the complex vector $\eta \in  \mathbb{C}^{V\times 1}$.
Now, the reweighting factors in Eqs.~\eqref{eq:WND}--\eqref{eq:Wll} can be estimated by using the integral identity in Eq.~\eqref{eq:A}, e.g.~for $R_{\rm ND}(\tilde{m},\tilde{m}',\mu_\delta,\mu_\delta') $ in Eq.~\eqref{eq:WND} follows
\begin{equation}
  R_{\rm ND} = \frac{1}{N_{\rm st}}\sum_{i=1}^{N_{\rm st}} \frac{e^{-\chi_i^\dagger D_{\rm ND}(\tilde{m},\mu_\delta) D_{\rm ND}^{-1}(\tilde{m}',\mu_\delta') \chi_i}}{e^{-\chi_i^\dagger\chi_i}} + \mathcal{O}(N_{\rm st}^{-1/2}).
\end{equation}
In order to reduce stochastic noise, we further split up the determinant ratios in Eq.~\eqref{eq:Wdet} by introducing three intermediate steps with 
\begin{equation}
 R_w = \prod_{j=0}^3 R_{w,j}(\tilde{m}_{j},\tilde{m}_{j+1},\mu_{\delta,j},\mu_{\delta,j+1})\,,
\end{equation}
where $\tilde{m}_j= ( (4-j) \tilde{m} + j \tilde{m}')/ 4 $ and $\mu_{\delta,j} = ( (4-j) \mu_\delta + j \mu_\delta')/ 4$, and using $N_{st}=16$ sufficiently suppresses the stochastic fluctuations.
Now, observables $\mathcal{O}(\tilde{m}',\mu_\delta')$, like correlations functions, can be evaluated at the new parameter set ${\{\tilde{m}',\mu_\delta'\}}$ via
\begin{equation}
 \langle \mathcal{O}(\tilde{m}',\mu_\delta') \rangle_{\{\tilde{m}',\mu_\delta'\}} = \frac{\langle \mathcal{O}(\tilde{m}',\mu_\delta') R_w \rangle_{\{\tilde{m},\mu_\delta\}}} {\langle R_w \rangle_{\{\tilde{m},\mu_\delta\}}}\,,
\end{equation}
using the ensemble generated at the old parameter set ${\{\tilde{m},\mu_\delta\}}$, where $\langle \cdots \rangle$ is the the ensemble average.
Splitting up the reweighting factor $R_w$ in several steps enables us to calculate the pion mass for the intermediate three steps, as shown in Fig.~\ref{fig:MpiA60}.
After reweighting in steps, we reach the value of $m_s$ of the A60s ensemble and a value of $M_\pi$ which agrees with the one extracted by a direct analysis of A60s. 

Having demonstrated that reweighting the strange-quark mass while maintaining maximal twist works, we can apply it to the cA211.30.32 ensemble.
For this ensemble, tuning to maximal twist is done at fixed strange- and charm-quark mass. Since we do not have a second ensemble at a different strange-quark mass tuned to maximal twist to know the target parameters, we now proceed in two steps:
we first change the strange-quark mass and then
re-adjust the bare light mass parameter $\tilde{m}$ to maximal twist. This is done while keeping the charm-quark mass constant.
The change in the strange-quark mass $m_s\rightarrow m_s + \delta m_s = m_s'$ can be corrected by the reweighting factor
$ R_1 = R_{\rm ND}(m_s,m_s')$.
Tuning to maximal twist at $m_s'$ requires the change in the bare mass parameter $\tilde{m} \rightarrow \tilde{m}+\delta \tilde{m} = \tilde{m}'$, which can be taken into account via the reweighting factor $R_2 =  R_\ell(\tilde{m},\tilde{m}') R_{\rm ND}(\tilde{m},\tilde{m}')$.

\begin{figure}
  \centering
  \includegraphics[width=0.48\textwidth]{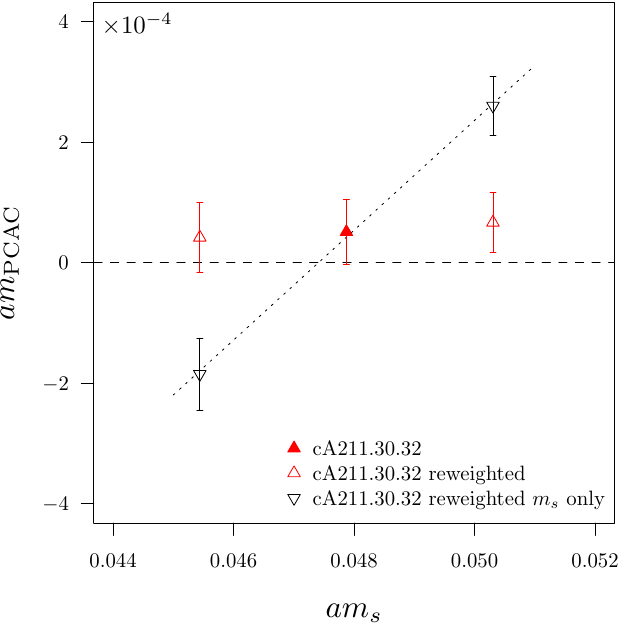}
  \caption{PCAC mass as a function of $m_s$ for cA211.30.32, without
  reweighting (red closed triangle), with reweighting but without
  tuning to maximal twist (black open triangles), and after tuning to
  maximal twist (red open triangles).}
  \label{fig:mpcac}
\end{figure}

For the cA211.30.32 ensemble, the stochastic fluctuations are suppressed by using $N_{\rm st}=64$ for each reweighting factor and performing a single step.
With the stochastic evaluation of Eq.~\eqref{eq:A}, the fluctuations are
\begin{align}
 \sigma^2_{\rm st} &=   \textrm{var}_\chi(R_\chi)/|R|^2, 
 \intertext{where} R_\chi &= \frac{e^{-\chi^\dagger A \chi}}{e^{-\chi^\dagger\chi}}
 \intertext{and the variance is given by} \textrm{var}_x(y) &= \sum^N_x \frac{y(x)^2}{N} - \left(\sum^N_x \frac{y(x)}{N}\right)^2.
\end{align}
The corresponding ratio matrix $A$ is, for the specific example of $ R_1$, given by $A=D_{\rm ND}(\tilde{m},\mu_\delta) D_{\rm ND}^{-1}(\tilde{m}',\mu_\delta')$.  
We find that the stochastic fluctuations are small compared to the statistical gauge fluctuations
\begin{equation}
 \sigma^2_{\rm ens} = \textrm{var}( R )/ \langle R \rangle^2 = \langle R^2 \rangle / \langle R \rangle^2 -1.
\end{equation}
For the two different steps, we obtain $\sigma^2_{\rm st}(R_1)/\sigma^2_{\rm ens}(R_1)\sim 0.02$ and $\sigma^2_{\rm st}(R_2)/\sigma^2_{\rm ens}(R_2)\sim 0.16$,
similar to what was obtained for clover Wilson fermions, see e.g.~\cite{Leder:2015fea,Finkenrath:2015ava}.

The parameters used for the cA211.30.32 ensemble for the different reweighting factors are listed in Table~\ref{tab:rew}. We vary the strange-quark mass by $\sim 5 \; \textrm{MeV}$. 
This leads to a significant shift of the PCAC mass (defined in Eq.~(6) of Ref.~\cite{Alexandrou:2018egz}), as shown in Fig.~\ref{fig:mpcac}. We then readjust by using reweighting in the bare quark mass parameters, following the tuning process 
outlined in Ref.~\cite{Alexandrou:2018egz}, which establishes a dependence of the PCAC mass on the bare mass parameter given by
\begin{equation}
 am_{\textrm{PCAC}}= 1.19(87) +  2.77(202) a{\tilde{m}}\,.
 \end{equation}
This leads to $\delta \tilde{m}$ given by 
\begin{equation}
 \delta \tilde{m}= \frac{\delta m_{\textrm{PCAC}}}{2.77}\, ,
\end{equation}
which indeed re-adjusts the PCAC mass to zero, as shown in Fig.~\ref{fig:mpcac} and listed in Table~\ref{tab:rew}.
After reweighting the ensemble cA211.30.32 by varying the
strange-quark mass by $\pm5\%$ around its original value and keeping a constant and maximal twist, 
the resulting values of $\beta_1$ and $\beta_2$ can be derived as listed in Table \ref{tab:beta12},
where correlations between the data points reduce the total statistical error.

\begin{table}
   \centering
   \begin{tabular}{@{\extracolsep{\fill}}lccccc}
     \hline\hline \\[-2.0ex]
      Ens.            &  $m_s$ &  $a \mu_{\sigma}$ & $a \mu_{\delta}$ & $a \tilde{m}$ & $10^{5} a m_{\textrm{PCAC}} $ \\
     \hline\hline \\[-2.0ex]
                    cA211 & $108.4$ & $0.1408$ & $0.1521$  & $0.43022$ & $5.0(53)$ \\ .30.32  \\
     \hline\hline      \\[-2.0ex]            
     Rew.            & $\Delta m_s$ &  $a \Delta \mu_{\sigma}$ & $a\Delta \mu_{\delta}$ & $a \Delta \tilde{m}$ &  $10^{5}a m_{\textrm{PCAC}}$ \\

     \hline\hline\\[-2.0ex]        $R_1$ & $+5.2$ & $+5.9 \cdot 10^{-4}$ & $ -7.6 \cdot 10^{-4}$  & -- & $+26.0(4.8)$  \\
                  $R_1$ & $-5.2$ & $-5.9\cdot 10^{-4}$ & $+7.5 \cdot 10^{-4}$  & --            & $-18.5(59)$ \\
                  \hline \\[-2.0ex] $R_2$ & $+5.2$ & -- & -- & $+7.2 \cdot 10^{-5}$ & $+6.6(50)$ \\
                  $R_2$ & $-5.2$ & -- & --  & $-7.2 \cdot 10^{-5}$                  & $+4.1(58)$ \\
     \hline\hline
   \end{tabular}
   \caption{Shift in the parameters employed by the different reweighting factors for the ensemble cA211.30.32. The strange-quark mass and the shifts are displayed in the $\overline{\textrm{MS}}$ scheme in MeV.
   In addition, the PCAC mass $m_{\rm PCAC}$ (defined in Eq.~(6) of Ref.~\cite{Alexandrou:2018egz}) is shown for each reweighting step.} 
\label{tab:rew}
\end{table}

\bibliography{bibliography}

\begin{thebibliography}{64}%
\makeatletter
\providecommand \@ifxundefined [1]{%
 \@ifx{#1\undefined}
}%
\providecommand \@ifnum [1]{%
 \ifnum #1\expandafter \@firstoftwo
 \else \expandafter \@secondoftwo
 \fi
}%
\providecommand \@ifx [1]{%
 \ifx #1\expandafter \@firstoftwo
 \else \expandafter \@secondoftwo
 \fi
}%
\providecommand \natexlab [1]{#1}%
\providecommand \enquote  [1]{``#1''}%
\providecommand \bibnamefont  [1]{#1}%
\providecommand \bibfnamefont [1]{#1}%
\providecommand \citenamefont [1]{#1}%
\providecommand \href@noop [0]{\@secondoftwo}%
\providecommand \href [0]{\begingroup \@sanitize@url \@href}%
\providecommand \@href[1]{\@@startlink{#1}\@@href}%
\providecommand \@@href[1]{\endgroup#1\@@endlink}%
\providecommand \@sanitize@url [0]{\catcode `\\12\catcode `\$12\catcode
  `\&12\catcode `\#12\catcode `\^12\catcode `\_12\catcode `\%12\relax}%
\providecommand \@@startlink[1]{}%
\providecommand \@@endlink[0]{}%
\providecommand \url  [0]{\begingroup\@sanitize@url \@url }%
\providecommand \@url [1]{\endgroup\@href {#1}{\urlprefix }}%
\providecommand \urlprefix  [0]{URL }%
\providecommand \Eprint [0]{\href }%
\providecommand \doibase [0]{https://doi.org/}%
\providecommand \selectlanguage [0]{\@gobble}%
\providecommand \bibinfo  [0]{\@secondoftwo}%
\providecommand \bibfield  [0]{\@secondoftwo}%
\providecommand \translation [1]{[#1]}%
\providecommand \BibitemOpen [0]{}%
\providecommand \bibitemStop [0]{}%
\providecommand \bibitemNoStop [0]{.\EOS\space}%
\providecommand \EOS [0]{\spacefactor3000\relax}%
\providecommand \BibitemShut  [1]{\csname bibitem#1\endcsname}%
\let\auto@bib@innerbib\@empty
\bibitem [{\citenamefont {Crewther}\ \emph {et~al.}(1979)\citenamefont
  {Crewther}, \citenamefont {Vecchia}, \citenamefont {Veneziano},\ and\
  \citenamefont {Witten}}]{Crewther:1979}%
  \BibitemOpen
  \bibfield  {author} {\bibinfo {author} {\bibfnamefont {R.~J.}\ \bibnamefont
  {Crewther}}, \bibinfo {author} {\bibfnamefont {P.~D.}\ \bibnamefont
  {Vecchia}}, \bibinfo {author} {\bibfnamefont {G.}~\bibnamefont {Veneziano}},\
  and\ \bibinfo {author} {\bibfnamefont {E.}~\bibnamefont {Witten}},\
  }\bibfield  {title} {\bibinfo {title} {Chiral estimate of the electric dipole
  moment of the neutron in quantum chromodynamics},\ }\href
  {https://doi.org/https://doi.org/10.1016/0370-2693(79)90128-X} {\bibfield
  {journal} {\bibinfo  {journal} {Phys. Lett. B}\ }\textbf {\bibinfo {volume}
  {88}},\ \bibinfo {pages} {123} (\bibinfo {year} {1979})}\BibitemShut
  {NoStop}%
\bibitem [{\citenamefont {Pendlebury}\ \emph {et~al.}(2015)\citenamefont
  {Pendlebury} \emph {et~al.}}]{Afach:2015sja}%
  \BibitemOpen
  \bibfield  {author} {\bibinfo {author} {\bibfnamefont {J.~M.}\ \bibnamefont
  {Pendlebury}} \emph {et~al.},\ }\bibfield  {title} {\bibinfo {title}
  {{Revised experimental upper limit on the electric dipole moment of the
  neutron}},\ }\href {https://doi.org/10.1103/PhysRevD.92.092003} {\bibfield
  {journal} {\bibinfo  {journal} {Phys. Rev. D}\ }\textbf {\bibinfo {volume}
  {92}},\ \bibinfo {pages} {092003} (\bibinfo {year} {2015})},\ \Eprint
  {https://arxiv.org/abs/1509.04411} {arXiv:1509.04411 [hep-ex]} \BibitemShut
  {NoStop}%
\bibitem [{\citenamefont {Abel}\ \emph {et~al.}(2020)\citenamefont {Abel} \emph
  {et~al.}}]{Abel:2020}%
  \BibitemOpen
  \bibfield  {author} {\bibinfo {author} {\bibfnamefont {C.}~\bibnamefont
  {Abel}} \emph {et~al.},\ }\bibfield  {title} {\bibinfo {title} {Measurement
  of the permanent electric dipole moment of the neutron},\ }\href
  {http://dx.doi.org/10.1103/PhysRevLett.124.081803} {\bibfield  {journal}
  {\bibinfo  {journal} {Phys. Rev. Lett.}\ }\textbf {\bibinfo {volume} {124}}
  (\bibinfo {year} {2020})}\BibitemShut {NoStop}%
\bibitem [{\citenamefont {Peccei}\ and\ \citenamefont
  {Quinn}(1977)}]{Peccei1977}%
  \BibitemOpen
  \bibfield  {author} {\bibinfo {author} {\bibfnamefont {R.~D.}\ \bibnamefont
  {Peccei}}\ and\ \bibinfo {author} {\bibfnamefont {H.~R.}\ \bibnamefont
  {Quinn}},\ }\bibfield  {title} {\bibinfo {title} {$\mathrm{{C}{P}}$
  {C}onservation in the {P}resence of {P}seudoparticles},\ }\href
  {https://doi.org/10.1103/PhysRevLett.38.1440} {\bibfield  {journal} {\bibinfo
   {journal} {Phys. Rev. Lett.}\ }\textbf {\bibinfo {volume} {38}},\ \bibinfo
  {pages} {1440} (\bibinfo {year} {1977})}\BibitemShut {NoStop}%
\bibitem [{\citenamefont {Weinberg}(1978)}]{Weinberg1977}%
  \BibitemOpen
  \bibfield  {author} {\bibinfo {author} {\bibfnamefont {S.}~\bibnamefont
  {Weinberg}},\ }\bibfield  {title} {\bibinfo {title} {{A} {N}ew {L}ight
  {B}oson?},\ }\href {https://doi.org/10.1103/PhysRevLett.40.223} {\bibfield
  {journal} {\bibinfo  {journal} {Phys. Rev. Lett.}\ }\textbf {\bibinfo
  {volume} {40}},\ \bibinfo {pages} {223} (\bibinfo {year} {1978})}\BibitemShut
  {NoStop}%
\bibitem [{\citenamefont {Wilczek}(1978)}]{Wilczek1978}%
  \BibitemOpen
  \bibfield  {author} {\bibinfo {author} {\bibfnamefont {F.}~\bibnamefont
  {Wilczek}},\ }\bibfield  {title} {\bibinfo {title} {{{P}roblem of {S}trong
  ${P}$ and ${T}$ {I}nvariance in the {P}resence of {I}nstantons}},\ }\href
  {https://doi.org/10.1103/PhysRevLett.40.279} {\bibfield  {journal} {\bibinfo
  {journal} {Phys. Rev. Lett.}\ }\textbf {\bibinfo {volume} {40}},\ \bibinfo
  {pages} {279} (\bibinfo {year} {1978})}\BibitemShut {NoStop}%
\bibitem [{\citenamefont {Georgi}\ and\ \citenamefont
  {McArthur}(1981)}]{Georgi1981}%
  \BibitemOpen
  \bibfield  {author} {\bibinfo {author} {\bibfnamefont {H.}~\bibnamefont
  {Georgi}}\ and\ \bibinfo {author} {\bibfnamefont {I.~N.}\ \bibnamefont
  {McArthur}},\ }\bibfield  {title} {\bibinfo {title} {{I}nstantons and the u
  quark mass},\ }\href@noop {} {\bibfield  {journal} {\bibinfo  {journal}
  {Harvard preprint HUTP-81/A011}\ } (\bibinfo {year} {1981})}\BibitemShut
  {NoStop}%
\bibitem [{\citenamefont {Kaplan}\ and\ \citenamefont
  {Manohar}(1986)}]{Kaplan1986}%
  \BibitemOpen
  \bibfield  {author} {\bibinfo {author} {\bibfnamefont {D.~B.}\ \bibnamefont
  {Kaplan}}\ and\ \bibinfo {author} {\bibfnamefont {A.~V.}\ \bibnamefont
  {Manohar}},\ }\bibfield  {title} {\bibinfo {title} {{C}urrent-{M}ass {R}atios
  of the {L}ight {Q}uarks},\ }\href
  {https://doi.org/10.1103/PhysRevLett.56.2004} {\bibfield  {journal} {\bibinfo
   {journal} {Phys. Rev. Lett.}\ }\textbf {\bibinfo {volume} {56}},\ \bibinfo
  {pages} {2004} (\bibinfo {year} {1986})}\BibitemShut {NoStop}%
\bibitem [{\citenamefont {Choi}\ \emph {et~al.}(1988)\citenamefont {Choi},
  \citenamefont {Kim},\ and\ \citenamefont {Sze}}]{Choi1988}%
  \BibitemOpen
  \bibfield  {author} {\bibinfo {author} {\bibfnamefont {K.}~\bibnamefont
  {Choi}}, \bibinfo {author} {\bibfnamefont {C.~W.}\ \bibnamefont {Kim}},\ and\
  \bibinfo {author} {\bibfnamefont {W.~K.}\ \bibnamefont {Sze}},\ }\bibfield
  {title} {\bibinfo {title} {{M}ass {R}enormalization by {I}nstantons and the
  {S}trong $\mathrm{{C}{P}}$ {P}roblem},\ }\href
  {https://doi.org/10.1103/PhysRevLett.61.794} {\bibfield  {journal} {\bibinfo
  {journal} {Phys. Rev. Lett.}\ }\textbf {\bibinfo {volume} {61}},\ \bibinfo
  {pages} {794} (\bibinfo {year} {1988})}\BibitemShut {NoStop}%
\bibitem [{\citenamefont {Banks}\ \emph {et~al.}(1994)\citenamefont {Banks},
  \citenamefont {Nir},\ and\ \citenamefont {Seiberg}}]{Banks1994}%
  \BibitemOpen
  \bibfield  {author} {\bibinfo {author} {\bibfnamefont {T.}~\bibnamefont
  {Banks}}, \bibinfo {author} {\bibfnamefont {Y.}~\bibnamefont {Nir}},\ and\
  \bibinfo {author} {\bibfnamefont {N.}~\bibnamefont {Seiberg}},\ }\bibfield
  {title} {\bibinfo {title} {{{M}issing (up) mass, accidental anomalous
  symmetries, and the strong {C}{P} problem}},\ }in\ \href@noop {} {\emph
  {\bibinfo {booktitle} {Yukawa couplings and the origins of mass. Proceedings,
  2nd IFT Workshop, Gainesville, USA, February 11-13, 1994}}}\ (\bibinfo {year}
  {1994})\ pp.\ \bibinfo {pages} {26--41},\ \Eprint
  {https://arxiv.org/abs/hep-ph/9403203} {arXiv:hep-ph/9403203 [hep-ph]}
  \BibitemShut {NoStop}%
\bibitem [{\citenamefont {Leurer}\ \emph {et~al.}(1993)\citenamefont {Leurer},
  \citenamefont {Nir},\ and\ \citenamefont {Seiberg}}]{Leurer1992}%
  \BibitemOpen
  \bibfield  {author} {\bibinfo {author} {\bibfnamefont {M.}~\bibnamefont
  {Leurer}}, \bibinfo {author} {\bibfnamefont {Y.}~\bibnamefont {Nir}},\ and\
  \bibinfo {author} {\bibfnamefont {N.}~\bibnamefont {Seiberg}},\ }\bibfield
  {title} {\bibinfo {title} {{Mass matrix models}},\ }\href
  {https://doi.org/10.1016/0550-3213(93)90112-3} {\bibfield  {journal}
  {\bibinfo  {journal} {Nucl. Phys. B}\ }\textbf {\bibinfo {volume} {398}},\
  \bibinfo {pages} {319} (\bibinfo {year} {1993})},\ \Eprint
  {https://arxiv.org/abs/hep-ph/9212278} {arXiv:hep-ph/9212278 [hep-ph]}
  \BibitemShut {NoStop}%
\bibitem [{\citenamefont {Leurer}\ \emph {et~al.}(1994)\citenamefont {Leurer},
  \citenamefont {Nir},\ and\ \citenamefont {Seiberg}}]{Leurer1993}%
  \BibitemOpen
  \bibfield  {author} {\bibinfo {author} {\bibfnamefont {M.}~\bibnamefont
  {Leurer}}, \bibinfo {author} {\bibfnamefont {Y.}~\bibnamefont {Nir}},\ and\
  \bibinfo {author} {\bibfnamefont {N.}~\bibnamefont {Seiberg}},\ }\bibfield
  {title} {\bibinfo {title} {{Mass matrix models: The Sequel}},\ }\href
  {https://doi.org/10.1016/0550-3213(94)90074-4} {\bibfield  {journal}
  {\bibinfo  {journal} {Nucl. Phys. B}\ }\textbf {\bibinfo {volume} {420}},\
  \bibinfo {pages} {468} (\bibinfo {year} {1994})},\ \Eprint
  {https://arxiv.org/abs/hep-ph/9310320} {arXiv:hep-ph/9310320 [hep-ph]}
  \BibitemShut {NoStop}%
\bibitem [{\citenamefont {Nelson}\ and\ \citenamefont
  {Strassler}(1997)}]{Nelson1996}%
  \BibitemOpen
  \bibfield  {author} {\bibinfo {author} {\bibfnamefont {A.~E.}\ \bibnamefont
  {Nelson}}\ and\ \bibinfo {author} {\bibfnamefont {M.~J.}\ \bibnamefont
  {Strassler}},\ }\bibfield  {title} {\bibinfo {title} {{A Realistic
  supersymmetric model with composite quarks}},\ }\href
  {https://doi.org/10.1103/PhysRevD.56.4226} {\bibfield  {journal} {\bibinfo
  {journal} {Phys. Rev. D}\ }\textbf {\bibinfo {volume} {56}},\ \bibinfo
  {pages} {4226} (\bibinfo {year} {1997})},\ \Eprint
  {https://arxiv.org/abs/hep-ph/9607362} {arXiv:hep-ph/9607362 [hep-ph]}
  \BibitemShut {NoStop}%
\bibitem [{\citenamefont {Kaplan}\ \emph {et~al.}(1999)\citenamefont {Kaplan},
  \citenamefont {Lepeintre}, \citenamefont {Masiero}, \citenamefont {Nelson},\
  and\ \citenamefont {Riotto}}]{Kaplan1998}%
  \BibitemOpen
  \bibfield  {author} {\bibinfo {author} {\bibfnamefont {D.~E.}\ \bibnamefont
  {Kaplan}}, \bibinfo {author} {\bibfnamefont {F.}~\bibnamefont {Lepeintre}},
  \bibinfo {author} {\bibfnamefont {A.}~\bibnamefont {Masiero}}, \bibinfo
  {author} {\bibfnamefont {A.~E.}\ \bibnamefont {Nelson}},\ and\ \bibinfo
  {author} {\bibfnamefont {A.}~\bibnamefont {Riotto}},\ }\bibfield  {title}
  {\bibinfo {title} {{Fermion masses and gauge mediated supersymmetry breaking
  from a single U(1)}},\ }\href {https://doi.org/10.1103/PhysRevD.60.055003}
  {\bibfield  {journal} {\bibinfo  {journal} {Phys. Rev. D}\ }\textbf {\bibinfo
  {volume} {60}},\ \bibinfo {pages} {055003} (\bibinfo {year} {1999})},\
  \Eprint {https://arxiv.org/abs/hep-ph/9806430} {arXiv:hep-ph/9806430
  [hep-ph]} \BibitemShut {NoStop}%
\bibitem [{\citenamefont {Nelson}(1984)}]{Nelson1983}%
  \BibitemOpen
  \bibfield  {author} {\bibinfo {author} {\bibfnamefont {A.~E.}\ \bibnamefont
  {Nelson}},\ }\bibfield  {title} {\bibinfo {title} {{Naturally Weak CP
  Violation}},\ }\href {https://doi.org/10.1016/0370-2693(84)92025-2}
  {\bibfield  {journal} {\bibinfo  {journal} {Phys. Lett.}\ }\textbf {\bibinfo
  {volume} {136B}},\ \bibinfo {pages} {387} (\bibinfo {year}
  {1984})}\BibitemShut {NoStop}%
\bibitem [{\citenamefont {Barr}(1984)}]{Barr1984}%
  \BibitemOpen
  \bibfield  {author} {\bibinfo {author} {\bibfnamefont {S.~M.}\ \bibnamefont
  {Barr}},\ }\bibfield  {title} {\bibinfo {title} {{Solving the Strong
  $\mathrm{CP}$ Problem without the Peccei-Quinn Symmetry}},\ }\href
  {https://doi.org/10.1103/PhysRevLett.53.329} {\bibfield  {journal} {\bibinfo
  {journal} {Phys. Rev. Lett.}\ }\textbf {\bibinfo {volume} {53}},\ \bibinfo
  {pages} {329} (\bibinfo {year} {1984})}\BibitemShut {NoStop}%
\bibitem [{\citenamefont {Dine}\ and\ \citenamefont
  {Draper}(2015)}]{Dine:2015jga}%
  \BibitemOpen
  \bibfield  {author} {\bibinfo {author} {\bibfnamefont {M.}~\bibnamefont
  {Dine}}\ and\ \bibinfo {author} {\bibfnamefont {P.}~\bibnamefont {Draper}},\
  }\bibfield  {title} {\bibinfo {title} {{Challenges for the Nelson-Barr
  Mechanism}},\ }\href {https://doi.org/10.1007/JHEP08(2015)132} {\bibfield
  {journal} {\bibinfo  {journal} {J. High Energy Phys.{}}\ }\textbf {\bibinfo
  {volume} {08}},\ \bibinfo {pages} {132} (\bibinfo {year} {2015})},\ \Eprint
  {https://arxiv.org/abs/1506.05433} {arXiv:1506.05433 [hep-ph]} \BibitemShut
  {NoStop}%
\bibitem [{\citenamefont {Bazavov}\ \emph
  {et~al.}(2018{\natexlab{a}})\citenamefont {Bazavov} \emph
  {et~al.}}]{Bazavov:2017lyh}%
  \BibitemOpen
  \bibfield  {author} {\bibinfo {author} {\bibfnamefont {A.}~\bibnamefont
  {Bazavov}} \emph {et~al.} (\bibinfo {collaboration} {Fermilab Lattice,
  MILC}),\ }\bibfield  {title} {\bibinfo {title} {{$B$- and $D$-meson leptonic
  decay constants from four-flavor lattice QCD}},\ }\href
  {https://doi.org/10.1103/PhysRevD.98.074512} {\bibfield  {journal} {\bibinfo
  {journal} {Phys. Rev. D}\ }\textbf {\bibinfo {volume} {98}},\ \bibinfo
  {pages} {074512} (\bibinfo {year} {2018}{\natexlab{a}})},\ \Eprint
  {https://arxiv.org/abs/1712.09262} {arXiv:1712.09262 [hep-lat]} \BibitemShut
  {NoStop}%
\bibitem [{\citenamefont {Bazavov}\ \emph
  {et~al.}(2018{\natexlab{b}})\citenamefont {Bazavov} \emph
  {et~al.}}]{Bazavov:2018omf}%
  \BibitemOpen
  \bibfield  {author} {\bibinfo {author} {\bibfnamefont {A.}~\bibnamefont
  {Bazavov}} \emph {et~al.} (\bibinfo {collaboration} {Fermilab Lattice, MILC,
  TUMQCD}),\ }\bibfield  {title} {\bibinfo {title} {{Up-, down-, strange-,
  charm-, and bottom-quark masses from four-flavor lattice QCD}},\ }\href
  {https://doi.org/10.1103/PhysRevD.98.054517} {\bibfield  {journal} {\bibinfo
  {journal} {Phys. Rev. D}\ }\textbf {\bibinfo {volume} {98}},\ \bibinfo
  {pages} {054517} (\bibinfo {year} {2018}{\natexlab{b}})},\ \Eprint
  {https://arxiv.org/abs/1802.04248} {arXiv:1802.04248 [hep-lat]} \BibitemShut
  {NoStop}%
\bibitem [{\citenamefont {Aoki}\ \emph {et~al.}(2020)\citenamefont {Aoki} \emph
  {et~al.}}]{Aoki:2019cca}%
  \BibitemOpen
  \bibfield  {author} {\bibinfo {author} {\bibfnamefont {S.}~\bibnamefont
  {Aoki}} \emph {et~al.} (\bibinfo {collaboration} {Flavour Lattice Averaging
  Group}),\ }\bibfield  {title} {\bibinfo {title} {{FLAG Review 2019: Flavour
  Lattice Averaging Group (FLAG)}},\ }\href
  {https://link.springer.com/article/10.1140/epjc/s10052-019-7354-7} {\bibfield
   {journal} {\bibinfo  {journal} {Eur. Phys. J. C}\ }\textbf {\bibinfo
  {volume} {80}},\ \bibinfo {pages} {113} (\bibinfo {year} {2020})},\ \Eprint
  {https://arxiv.org/abs/1902.08191} {arXiv:1902.08191 [hep-lat]} \BibitemShut
  {NoStop}%
\bibitem [{\citenamefont {Cohen}\ \emph {et~al.}(1999)\citenamefont {Cohen},
  \citenamefont {Kaplan},\ and\ \citenamefont {Nelson}}]{Cohen1999}%
  \BibitemOpen
  \bibfield  {author} {\bibinfo {author} {\bibfnamefont {A.~G.}\ \bibnamefont
  {Cohen}}, \bibinfo {author} {\bibfnamefont {D.~B.}\ \bibnamefont {Kaplan}},\
  and\ \bibinfo {author} {\bibfnamefont {A.~E.}\ \bibnamefont {Nelson}},\
  }\bibfield  {title} {\bibinfo {title} {{Testing m(u)=0 on the lattice}},\
  }\href {https://doi.org/10.1088/1126-6708/1999/11/027} {\bibfield  {journal}
  {\bibinfo  {journal} {J. High Energy Phys.{}}\ }\textbf {\bibinfo {volume}
  {11}},\ \bibinfo {pages} {027} (\bibinfo {year} {1999})},\ \Eprint
  {https://arxiv.org/abs/hep-lat/9909091} {arXiv:hep-lat/9909091 [hep-lat]}
  \BibitemShut {NoStop}%
\bibitem [{\citenamefont {Dine}\ \emph {et~al.}(2015)\citenamefont {Dine},
  \citenamefont {Draper},\ and\ \citenamefont {Festuccia}}]{Dine:2014dga}%
  \BibitemOpen
  \bibfield  {author} {\bibinfo {author} {\bibfnamefont {M.}~\bibnamefont
  {Dine}}, \bibinfo {author} {\bibfnamefont {P.}~\bibnamefont {Draper}},\ and\
  \bibinfo {author} {\bibfnamefont {G.}~\bibnamefont {Festuccia}},\ }\bibfield
  {title} {\bibinfo {title} {{Instanton Effects in Three Flavor QCD}},\ }\href
  {https://doi.org/10.1103/PhysRevD.92.054004} {\bibfield  {journal} {\bibinfo
  {journal} {Phys. Rev. D}\ }\textbf {\bibinfo {volume} {92}},\ \bibinfo
  {pages} {054004} (\bibinfo {year} {2015})},\ \Eprint
  {https://arxiv.org/abs/1410.8505} {arXiv:1410.8505 [hep-ph]} \BibitemShut
  {NoStop}%
\bibitem [{Note1()}]{Note1}%
  \BibitemOpen
  \bibinfo {note} {Note that, in this work, we can use larger than physical
  quark masses due to reasons discussed below Eq.~\protect \textup {\hbox
  {\mathsurround \z@ \protect \normalfont (\ignorespaces \ref {eq:beta}\unskip
  \@@italiccorr )}}.}\BibitemShut {Stop}%
\bibitem [{\citenamefont {Bazavov}\ \emph {et~al.}(2013)\citenamefont {Bazavov}
  \emph {et~al.}}]{Bazavov:2012xda}%
  \BibitemOpen
  \bibfield  {author} {\bibinfo {author} {\bibfnamefont {A.}~\bibnamefont
  {Bazavov}} \emph {et~al.} (\bibinfo {collaboration} {MILC}),\ }\bibfield
  {title} {\bibinfo {title} {{Lattice QCD Ensembles with Four Flavors of Highly
  Improved Staggered Quarks}},\ }\href
  {https://doi.org/10.1103/PhysRevD.87.054505} {\bibfield  {journal} {\bibinfo
  {journal} {Phys. Rev. D}\ }\textbf {\bibinfo {volume} {87}},\ \bibinfo
  {pages} {054505} (\bibinfo {year} {2013})},\ \Eprint
  {https://arxiv.org/abs/1212.4768} {arXiv:1212.4768 [hep-lat]} \BibitemShut
  {NoStop}%
\bibitem [{\citenamefont {Alexandrou}\ \emph {et~al.}(2018)\citenamefont
  {Alexandrou} \emph {et~al.}}]{Alexandrou:2018egz}%
  \BibitemOpen
  \bibfield  {author} {\bibinfo {author} {\bibfnamefont {C.}~\bibnamefont
  {Alexandrou}} \emph {et~al.},\ }\bibfield  {title} {\bibinfo {title}
  {{Simulating twisted mass fermions at physical light, strange and charm quark
  masses}},\ }\href {https://doi.org/10.1103/PhysRevD.98.054518} {\bibfield
  {journal} {\bibinfo  {journal} {Phys. Rev. D}\ }\textbf {\bibinfo {volume}
  {98}},\ \bibinfo {pages} {054518} (\bibinfo {year} {2018})},\ \Eprint
  {https://arxiv.org/abs/1807.00495} {arXiv:1807.00495 [hep-lat]} \BibitemShut
  {NoStop}%
\bibitem [{\citenamefont {Alexandrou}\ \emph {et~al.}(2017)\citenamefont
  {Alexandrou}, \citenamefont {Athenodorou}, \citenamefont {Cichy},
  \citenamefont {Dromard}, \citenamefont {Garcia-Ramos}, \citenamefont
  {Jansen}, \citenamefont {Wenger},\ and\ \citenamefont
  {Zimmermann}}]{Alexandrou:2017hqw}%
  \BibitemOpen
  \bibfield  {author} {\bibinfo {author} {\bibfnamefont {C.}~\bibnamefont
  {Alexandrou}}, \bibinfo {author} {\bibfnamefont {A.}~\bibnamefont
  {Athenodorou}}, \bibinfo {author} {\bibfnamefont {K.}~\bibnamefont {Cichy}},
  \bibinfo {author} {\bibfnamefont {A.}~\bibnamefont {Dromard}}, \bibinfo
  {author} {\bibfnamefont {E.}~\bibnamefont {Garcia-Ramos}}, \bibinfo {author}
  {\bibfnamefont {K.}~\bibnamefont {Jansen}}, \bibinfo {author} {\bibfnamefont
  {U.}~\bibnamefont {Wenger}},\ and\ \bibinfo {author} {\bibfnamefont
  {F.}~\bibnamefont {Zimmermann}},\ }\bibfield  {title} {\bibinfo {title}
  {{Comparison of topological charge definitions in Lattice QCD}},\ }\href@noop
  {} {\  (\bibinfo {year} {2017})},\ \Eprint {https://arxiv.org/abs/1708.00696}
  {arXiv:1708.00696 [hep-lat]} \BibitemShut {NoStop}%
\bibitem [{\citenamefont {Nelson}\ \emph {et~al.}(2003)\citenamefont {Nelson},
  \citenamefont {Fleming},\ and\ \citenamefont {Kilcup}}]{Nelson2003}%
  \BibitemOpen
  \bibfield  {author} {\bibinfo {author} {\bibfnamefont {D.~R.}\ \bibnamefont
  {Nelson}}, \bibinfo {author} {\bibfnamefont {G.~T.}\ \bibnamefont
  {Fleming}},\ and\ \bibinfo {author} {\bibfnamefont {G.~W.}\ \bibnamefont
  {Kilcup}},\ }\bibfield  {title} {\bibinfo {title} {{Is strong CP due to a
  massless up quark?}},\ }\href {https://doi.org/10.1103/PhysRevLett.90.021601}
  {\bibfield  {journal} {\bibinfo  {journal} {Phys. Rev. Lett.}\ }\textbf
  {\bibinfo {volume} {90}},\ \bibinfo {pages} {021601} (\bibinfo {year}
  {2003})},\ \Eprint {https://arxiv.org/abs/hep-lat/0112029}
  {arXiv:hep-lat/0112029 [hep-lat]} \BibitemShut {NoStop}%
\bibitem [{\citenamefont {Cline}(1989)}]{Cline1989}%
  \BibitemOpen
  \bibfield  {author} {\bibinfo {author} {\bibfnamefont {J.~M.}\ \bibnamefont
  {Cline}},\ }\bibfield  {title} {\bibinfo {title} {{Can
  ${\ensuremath{\theta}}_{\mathrm{QCD}}=\mathrm{\ensuremath{\pi}}$?}},\ }\href
  {https://doi.org/10.1103/PhysRevLett.63.1338} {\bibfield  {journal} {\bibinfo
   {journal} {Phys. Rev. Lett.}\ }\textbf {\bibinfo {volume} {63}},\ \bibinfo
  {pages} {1338} (\bibinfo {year} {1989})}\BibitemShut {NoStop}%
\bibitem [{\citenamefont {Dragos}\ \emph {et~al.}(2019)\citenamefont {Dragos},
  \citenamefont {Luu}, \citenamefont {Shindler}, \citenamefont {de~Vries},\
  and\ \citenamefont {Yousif}}]{Dragos:2019oxn}%
  \BibitemOpen
  \bibfield  {author} {\bibinfo {author} {\bibfnamefont {J.}~\bibnamefont
  {Dragos}}, \bibinfo {author} {\bibfnamefont {T.}~\bibnamefont {Luu}},
  \bibinfo {author} {\bibfnamefont {A.}~\bibnamefont {Shindler}}, \bibinfo
  {author} {\bibfnamefont {J.}~\bibnamefont {de~Vries}},\ and\ \bibinfo
  {author} {\bibfnamefont {A.}~\bibnamefont {Yousif}},\ }\bibfield  {title}
  {\bibinfo {title} {{Confirming the Existence of the Strong CP Problem in
  Lattice QCD with the Gradient Flow}},\ }\href@noop {} {\  (\bibinfo {year}
  {2019})},\ \Eprint {https://arxiv.org/abs/1902.03254} {arXiv:1902.03254
  [hep-lat]} \BibitemShut {NoStop}%
\bibitem [{Note2()}]{Note2}%
  \BibitemOpen
  \bibinfo {note} {Note that $\beta _1$ is dimensionless as in Ref.~\cite
  {Banks1994}, while the dimensionful $\beta _2$ translates to $\beta _2 /
  \Lambda _{\protect \rm \chi SB}$ in Ref.~\cite {Banks1994}.}\BibitemShut
  {Stop}%
\bibitem [{\citenamefont {Aoki}\ \emph {et~al.}(2014)\citenamefont {Aoki} \emph
  {et~al.}}]{Aoki2013}%
  \BibitemOpen
  \bibfield  {author} {\bibinfo {author} {\bibfnamefont {S.}~\bibnamefont
  {Aoki}} \emph {et~al.},\ }\bibfield  {title} {\bibinfo {title} {{Review of
  lattice results concerning low-energy particle physics}},\ }\href
  {https://doi.org/10.1140/epjc/s10052-014-2890-7} {\bibfield  {journal}
  {\bibinfo  {journal} {Eur. Phys. J. C}\ }\textbf {\bibinfo {volume} {74}},\
  \bibinfo {pages} {2890} (\bibinfo {year} {2014})},\ \Eprint
  {https://arxiv.org/abs/1310.8555} {arXiv:1310.8555 [hep-lat]} \BibitemShut
  {NoStop}%
\bibitem [{\citenamefont {Iwasaki}(1985)}]{Iwasaki:1985we}%
  \BibitemOpen
  \bibfield  {author} {\bibinfo {author} {\bibfnamefont {Y.}~\bibnamefont
  {Iwasaki}},\ }\bibfield  {title} {\bibinfo {title} {{Renormalization Group
  Analysis of Lattice Theories and Improved Lattice Action: Two-Dimensional
  Nonlinear O(N) Sigma Model}},\ }\href
  {https://doi.org/10.1016/0550-3213(85)90606-6} {\bibfield  {journal}
  {\bibinfo  {journal} {Nucl. Phys. B}\ }\textbf {\bibinfo {volume} {258}},\
  \bibinfo {pages} {141} (\bibinfo {year} {1985})}\BibitemShut {NoStop}%
\bibitem [{\citenamefont {Frezzotti}\ \emph {et~al.}(2001)\citenamefont
  {Frezzotti}, \citenamefont {Grassi}, \citenamefont {Sint},\ and\
  \citenamefont {Weisz}}]{Frezzotti:2000nk}%
  \BibitemOpen
  \bibfield  {author} {\bibinfo {author} {\bibfnamefont {R.}~\bibnamefont
  {Frezzotti}}, \bibinfo {author} {\bibfnamefont {P.~A.}\ \bibnamefont
  {Grassi}}, \bibinfo {author} {\bibfnamefont {S.}~\bibnamefont {Sint}},\ and\
  \bibinfo {author} {\bibfnamefont {P.}~\bibnamefont {Weisz}} (\bibinfo
  {collaboration} {Alpha}),\ }\bibfield  {title} {\bibinfo {title} {{Lattice
  QCD with a chirally twisted mass term}},\ }\href
  {https://iopscience.iop.org/article/10.1088/1126-6708/2001/08/058} {\bibfield
   {journal} {\bibinfo  {journal} {J. High Energy Phys.{}}\ }\textbf {\bibinfo
  {volume} {08}},\ \bibinfo {pages} {058} (\bibinfo {year} {2001})},\ \Eprint
  {https://arxiv.org/abs/hep-lat/0101001} {arXiv:hep-lat/0101001 [hep-lat]}
  \BibitemShut {NoStop}%
\bibitem [{\citenamefont {Frezzotti}\ and\ \citenamefont
  {Rossi}(2004{\natexlab{a}})}]{Frezzotti:2003xj}%
  \BibitemOpen
  \bibfield  {author} {\bibinfo {author} {\bibfnamefont {R.}~\bibnamefont
  {Frezzotti}}\ and\ \bibinfo {author} {\bibfnamefont {G.~C.}\ \bibnamefont
  {Rossi}},\ }\bibfield  {title} {\bibinfo {title} {{Twisted mass lattice QCD
  with mass nondegenerate quarks}},\ }\bibfield  {booktitle} {\emph {\bibinfo
  {booktitle} {{Lattice hadron physics. Proceedings, 2nd Topical Workshop, LHP
  2003, Cairns, Australia, July 22-30, 2003}}},\ }\href
  {https://doi.org/10.1016/S0920-5632(03)02477-0} {\bibfield  {journal}
  {\bibinfo  {journal} {Nucl. Phys. Proc. Suppl.}\ }\textbf {\bibinfo {volume}
  {128}},\ \bibinfo {pages} {193} (\bibinfo {year} {2004}{\natexlab{a}})},\
  \Eprint {https://arxiv.org/abs/hep-lat/0311008} {arXiv:hep-lat/0311008
  [hep-lat]} \BibitemShut {NoStop}%
\bibitem [{Sup()}]{SupplMat}%
  \BibitemOpen
  \href@noop {} {}\bibinfo {note} {See Supplemental Material for strange-quark
  reweighting towards maximal twist.}\BibitemShut {Stop}%
\bibitem [{\citenamefont {Baron}\ \emph {et~al.}(2010)\citenamefont {Baron}
  \emph {et~al.}}]{Baron:2010bv}%
  \BibitemOpen
  \bibfield  {author} {\bibinfo {author} {\bibfnamefont {R.}~\bibnamefont
  {Baron}} \emph {et~al.},\ }\bibfield  {title} {\bibinfo {title} {{Light
  hadrons from lattice QCD with light (u,d), strange and charm dynamical
  quarks}},\ }\href {https://doi.org/10.1007/JHEP06(2010)111} {\bibfield
  {journal} {\bibinfo  {journal} {J. High Energy Phys.{}}\ }\textbf {\bibinfo
  {volume} {06}},\ \bibinfo {pages} {111} (\bibinfo {year} {2010})},\ \Eprint
  {https://arxiv.org/abs/1004.5284} {arXiv:1004.5284 [hep-lat]} \BibitemShut
  {NoStop}%
\bibitem [{\citenamefont {Carrasco}\ \emph {et~al.}(2014)\citenamefont
  {Carrasco} \emph {et~al.}}]{Carrasco:2014cwa}%
  \BibitemOpen
  \bibfield  {author} {\bibinfo {author} {\bibfnamefont {N.}~\bibnamefont
  {Carrasco}} \emph {et~al.} (\bibinfo {collaboration} {ETM}),\ }\bibfield
  {title} {\bibinfo {title} {{Up, down, strange and charm quark masses with
  N$_f$ = 2+1+1 twisted mass lattice QCD}},\ }\href
  {https://doi.org/10.1016/j.nuclphysb.2014.07.025} {\bibfield  {journal}
  {\bibinfo  {journal} {Nucl. Phys. B}\ }\textbf {\bibinfo {volume} {887}},\
  \bibinfo {pages} {19} (\bibinfo {year} {2014})},\ \Eprint
  {https://arxiv.org/abs/1403.4504} {arXiv:1403.4504 [hep-lat]} \BibitemShut
  {NoStop}%
\bibitem [{\citenamefont {Frezzotti}\ and\ \citenamefont
  {Rossi}(2004{\natexlab{b}})}]{Frezzotti:2003ni}%
  \BibitemOpen
  \bibfield  {author} {\bibinfo {author} {\bibfnamefont {R.}~\bibnamefont
  {Frezzotti}}\ and\ \bibinfo {author} {\bibfnamefont {G.~C.}\ \bibnamefont
  {Rossi}},\ }\bibfield  {title} {\bibinfo {title} {Chirally improving {Wilson}
  fermions. {I}: {O(a)} improvement},\ }\href
  {https://iopscience.iop.org/article/10.1088/1126-6708/2004/08/007/meta}
  {\bibfield  {journal} {\bibinfo  {journal} {J. High Energy Phys.{}}\ }\textbf
  {\bibinfo {volume} {08}},\ \bibinfo {pages} {007} (\bibinfo {year}
  {2004}{\natexlab{b}})},\ \Eprint {https://arxiv.org/abs/hep-lat/0306014}
  {hep-lat/0306014} \BibitemShut {NoStop}%
\bibitem [{\citenamefont {Abdel-Rehim}\ \emph {et~al.}(2017)\citenamefont
  {Abdel-Rehim} \emph {et~al.}}]{Abdel-Rehim:2015pwa}%
  \BibitemOpen
  \bibfield  {author} {\bibinfo {author} {\bibfnamefont {A.}~\bibnamefont
  {Abdel-Rehim}} \emph {et~al.} (\bibinfo {collaboration} {ETM}),\ }\bibfield
  {title} {\bibinfo {title} {{First physics results at the physical pion mass
  from $N_f=2$ Wilson twisted mass fermions at maximal twist}},\ }\href
  {https://doi.org/10.1103/PhysRevD.95.094515} {\bibfield  {journal} {\bibinfo
  {journal} {Phys. Rev. D}\ }\textbf {\bibinfo {volume} {95}},\ \bibinfo
  {pages} {094515} (\bibinfo {year} {2017})},\ \Eprint
  {https://arxiv.org/abs/1507.05068} {arXiv:1507.05068 [hep-lat]} \BibitemShut
  {NoStop}%
\bibitem [{\citenamefont {Gasser}\ \emph {et~al.}(1988)\citenamefont {Gasser},
  \citenamefont {Sainio},\ and\ \citenamefont {Svarc}}]{Gasser:1987rb}%
  \BibitemOpen
  \bibfield  {author} {\bibinfo {author} {\bibfnamefont {J.}~\bibnamefont
  {Gasser}}, \bibinfo {author} {\bibfnamefont {M.~E.}\ \bibnamefont {Sainio}},\
  and\ \bibinfo {author} {\bibfnamefont {A.}~\bibnamefont {Svarc}},\ }\bibfield
   {title} {\bibinfo {title} {{Nucleons with Chiral Loops}},\ }\href
  {https://doi.org/10.1016/0550-3213(88)90108-3} {\bibfield  {journal}
  {\bibinfo  {journal} {Nucl. Phys. B}\ }\textbf {\bibinfo {volume} {307}},\
  \bibinfo {pages} {779} (\bibinfo {year} {1988})}\BibitemShut {NoStop}%
\bibitem [{\citenamefont {Tiburzi}\ and\ \citenamefont
  {Walker-Loud}(2008)}]{Tiburzi:2008bk}%
  \BibitemOpen
  \bibfield  {author} {\bibinfo {author} {\bibfnamefont {B.~C.}\ \bibnamefont
  {Tiburzi}}\ and\ \bibinfo {author} {\bibfnamefont {A.}~\bibnamefont
  {Walker-Loud}},\ }\bibfield  {title} {\bibinfo {title} {{Hyperons in Two
  Flavor Chiral Perturbation Theory}},\ }\href
  {https://doi.org/10.1016/j.physletb.2008.09.054} {\bibfield  {journal}
  {\bibinfo  {journal} {Phys. Lett. B}\ }\textbf {\bibinfo {volume} {669}},\
  \bibinfo {pages} {246} (\bibinfo {year} {2008})},\ \Eprint
  {https://arxiv.org/abs/0808.0482} {arXiv:0808.0482 [nucl-th]} \BibitemShut
  {NoStop}%
\bibitem [{\citenamefont {Alexandrou}\ \emph {et~al.}(2014)\citenamefont
  {Alexandrou}, \citenamefont {Drach}, \citenamefont {Jansen}, \citenamefont
  {Kallidonis},\ and\ \citenamefont {Koutsou}}]{Alexandrou:2014sha}%
  \BibitemOpen
  \bibfield  {author} {\bibinfo {author} {\bibfnamefont {C.}~\bibnamefont
  {Alexandrou}}, \bibinfo {author} {\bibfnamefont {V.}~\bibnamefont {Drach}},
  \bibinfo {author} {\bibfnamefont {K.}~\bibnamefont {Jansen}}, \bibinfo
  {author} {\bibfnamefont {C.}~\bibnamefont {Kallidonis}},\ and\ \bibinfo
  {author} {\bibfnamefont {G.}~\bibnamefont {Koutsou}},\ }\bibfield  {title}
  {\bibinfo {title} {{Baryon spectrum with $N_f=2+1+1$ twisted mass
  fermions}},\ }\href {https://doi.org/10.1103/PhysRevD.90.074501} {\bibfield
  {journal} {\bibinfo  {journal} {Phys. Rev. D}\ }\textbf {\bibinfo {volume}
  {90}},\ \bibinfo {pages} {074501} (\bibinfo {year} {2014})},\ \Eprint
  {https://arxiv.org/abs/1406.4310} {arXiv:1406.4310 [hep-lat]} \BibitemShut
  {NoStop}%
\bibitem [{\citenamefont {Bruno}\ \emph {et~al.}(2017)\citenamefont {Bruno},
  \citenamefont {Dalla~Brida}, \citenamefont {Fritzsch}, \citenamefont
  {Korzec}, \citenamefont {Ramos}, \citenamefont {Schaefer}, \citenamefont
  {Simma}, \citenamefont {Sint},\ and\ \citenamefont {Sommer}}]{Bruno:2017gxd}%
  \BibitemOpen
  \bibfield  {author} {\bibinfo {author} {\bibfnamefont {M.}~\bibnamefont
  {Bruno}}, \bibinfo {author} {\bibfnamefont {M.}~\bibnamefont {Dalla~Brida}},
  \bibinfo {author} {\bibfnamefont {P.}~\bibnamefont {Fritzsch}}, \bibinfo
  {author} {\bibfnamefont {T.}~\bibnamefont {Korzec}}, \bibinfo {author}
  {\bibfnamefont {A.}~\bibnamefont {Ramos}}, \bibinfo {author} {\bibfnamefont
  {S.}~\bibnamefont {Schaefer}}, \bibinfo {author} {\bibfnamefont
  {H.}~\bibnamefont {Simma}}, \bibinfo {author} {\bibfnamefont
  {S.}~\bibnamefont {Sint}},\ and\ \bibinfo {author} {\bibfnamefont
  {R.}~\bibnamefont {Sommer}} (\bibinfo {collaboration} {ALPHA}),\ }\bibfield
  {title} {\bibinfo {title} {{QCD Coupling from a Nonperturbative Determination
  of the Three-Flavor $\Lambda$ Parameter}},\ }\href
  {https://doi.org/10.1103/PhysRevLett.119.102001} {\bibfield  {journal}
  {\bibinfo  {journal} {Phys. Rev. Lett.}\ }\textbf {\bibinfo {volume} {119}},\
  \bibinfo {pages} {102001} (\bibinfo {year} {2017})},\ \Eprint
  {https://arxiv.org/abs/1706.03821} {arXiv:1706.03821 [hep-lat]} \BibitemShut
  {NoStop}%
\bibitem [{\citenamefont {Novikov}\ \emph {et~al.}(1981)\citenamefont
  {Novikov}, \citenamefont {Shifman}, \citenamefont {Vainshtein},\ and\
  \citenamefont {Zakharov}}]{Novikov1981}%
  \BibitemOpen
  \bibfield  {author} {\bibinfo {author} {\bibfnamefont {V.~A.}\ \bibnamefont
  {Novikov}}, \bibinfo {author} {\bibfnamefont {M.~A.}\ \bibnamefont
  {Shifman}}, \bibinfo {author} {\bibfnamefont {A.~I.}\ \bibnamefont
  {Vainshtein}},\ and\ \bibinfo {author} {\bibfnamefont {V.~I.}\ \bibnamefont
  {Zakharov}},\ }\bibfield  {title} {\bibinfo {title} {Are all hadrons
  alike?},\ }\href
  {https://doi.org/https://doi.org/10.1016/0550-3213(81)90303-5} {\bibfield
  {journal} {\bibinfo  {journal} {Nucl. Phys. B}\ }\textbf {\bibinfo {volume}
  {191}},\ \bibinfo {pages} {301} (\bibinfo {year} {1981})}\BibitemShut
  {NoStop}%
\bibitem [{\citenamefont {Gasser}\ and\ \citenamefont
  {Leutwyler}(1985)}]{Gasser1985}%
  \BibitemOpen
  \bibfield  {author} {\bibinfo {author} {\bibfnamefont {J.}~\bibnamefont
  {Gasser}}\ and\ \bibinfo {author} {\bibfnamefont {H.}~\bibnamefont
  {Leutwyler}},\ }\bibfield  {title} {\bibinfo {title} {Chiral perturbation
  theory: Expansions in the mass of the strange quark},\ }\href
  {https://doi.org/https://doi.org/10.1016/0550-3213(85)90492-4} {\bibfield
  {journal} {\bibinfo  {journal} {Nucl. Phys. B}\ }\textbf {\bibinfo {volume}
  {250}},\ \bibinfo {pages} {465} (\bibinfo {year} {1985})}\BibitemShut
  {NoStop}%
\bibitem [{\citenamefont {{J\"{u}lich Supercomputing Centre}}(2015)}]{juqueen}%
  \BibitemOpen
  \bibfield  {author} {\bibinfo {author} {\bibnamefont {{J\"{u}lich
  Supercomputing Centre}}},\ }\bibfield  {title} {\bibinfo {title} {{JUQUEEN:
  IBM Blue Gene/Q Supercomputer System at the J\"{u}lich Supercomputing
  Centre}},\ }\bibfield  {journal} {\bibinfo  {journal} {Journal of large-scale
  research facilities}\ }\textbf {\bibinfo {volume} {1}},\ \href
  {https://doi.org/10.17815/jlsrf-1-18} {10.17815/jlsrf-1-18} (\bibinfo {year}
  {2015})\BibitemShut {NoStop}%
\bibitem [{\citenamefont {{J\"{u}lich Supercomputing Centre}}(2018)}]{jureca}%
  \BibitemOpen
  \bibfield  {author} {\bibinfo {author} {\bibnamefont {{J\"{u}lich
  Supercomputing Centre}}},\ }\bibfield  {title} {\bibinfo {title} {{JURECA:
  Modular supercomputer at J\"{u}lich Supercomputing Centre}},\ }\bibfield
  {journal} {\bibinfo  {journal} {Journal of large-scale research facilities}\
  }\textbf {\bibinfo {volume} {4}},\ \href
  {https://doi.org/10.17815/jlsrf-4-121-1} {10.17815/jlsrf-4-121-1} (\bibinfo
  {year} {2018})\BibitemShut {NoStop}%
\bibitem [{\citenamefont {Jansen}\ and\ \citenamefont
  {Urbach}(2009)}]{Jansen:2009xp}%
  \BibitemOpen
  \bibfield  {author} {\bibinfo {author} {\bibfnamefont {K.}~\bibnamefont
  {Jansen}}\ and\ \bibinfo {author} {\bibfnamefont {C.}~\bibnamefont
  {Urbach}},\ }\bibfield  {title} {\bibinfo {title} {{tmLQCD: A Program suite
  to simulate Wilson Twisted mass Lattice QCD}},\ }\href
  {https://doi.org/10.1016/j.cpc.2009.05.016} {\bibfield  {journal} {\bibinfo
  {journal} {Comput. Phys. Commun.}\ }\textbf {\bibinfo {volume} {180}},\
  \bibinfo {pages} {2717} (\bibinfo {year} {2009})},\ \Eprint
  {https://arxiv.org/abs/0905.3331} {arXiv:0905.3331 [hep-lat]} \BibitemShut
  {NoStop}%
\bibitem [{\citenamefont {Deuzeman}\ \emph {et~al.}(2012)\citenamefont
  {Deuzeman}, \citenamefont {Reker},\ and\ \citenamefont
  {Urbach}}]{Deuzeman:2011wz}%
  \BibitemOpen
  \bibfield  {author} {\bibinfo {author} {\bibfnamefont {A.}~\bibnamefont
  {Deuzeman}}, \bibinfo {author} {\bibfnamefont {S.}~\bibnamefont {Reker}},\
  and\ \bibinfo {author} {\bibfnamefont {C.}~\bibnamefont {Urbach}} (\bibinfo
  {collaboration} {ETM}),\ }\bibfield  {title} {\bibinfo {title} {{Lemon: an
  MPI parallel I/O library for data encapsulation using LIME}},\ }\href
  {https://doi.org/10.1016/j.cpc.2012.01.016} {\bibfield  {journal} {\bibinfo
  {journal} {Comput. Phys. Commun.}\ }\textbf {\bibinfo {volume} {183}},\
  \bibinfo {pages} {1321} (\bibinfo {year} {2012})},\ \Eprint
  {https://arxiv.org/abs/1106.4177} {arXiv:1106.4177 [hep-lat]} \BibitemShut
  {NoStop}%
\bibitem [{\citenamefont {Frommer}\ \emph {et~al.}(2014)\citenamefont
  {Frommer}, \citenamefont {Kahl}, \citenamefont {Krieg}, \citenamefont
  {Leder},\ and\ \citenamefont {Rottmann}}]{Frommer:2013fsa}%
  \BibitemOpen
  \bibfield  {author} {\bibinfo {author} {\bibfnamefont {A.}~\bibnamefont
  {Frommer}}, \bibinfo {author} {\bibfnamefont {K.}~\bibnamefont {Kahl}},
  \bibinfo {author} {\bibfnamefont {S.}~\bibnamefont {Krieg}}, \bibinfo
  {author} {\bibfnamefont {B.}~\bibnamefont {Leder}},\ and\ \bibinfo {author}
  {\bibfnamefont {M.}~\bibnamefont {Rottmann}},\ }\bibfield  {title} {\bibinfo
  {title} {{Adaptive Aggregation Based Domain Decomposition Multigrid for the
  Lattice Wilson Dirac Operator}},\ }\href {https://doi.org/10.1137/130919507}
  {\bibfield  {journal} {\bibinfo  {journal} {SIAM J. Sci. Comput.}\ }\textbf
  {\bibinfo {volume} {36}},\ \bibinfo {pages} {A1581} (\bibinfo {year}
  {2014})},\ \Eprint {https://arxiv.org/abs/1303.1377} {arXiv:1303.1377
  [hep-lat]} \BibitemShut {NoStop}%
\bibitem [{\citenamefont {Alexandrou}\ \emph {et~al.}(2016)\citenamefont
  {Alexandrou}, \citenamefont {Bacchio}, \citenamefont {Finkenrath},
  \citenamefont {Frommer}, \citenamefont {Kahl},\ and\ \citenamefont
  {Rottmann}}]{Alexandrou:2016izb}%
  \BibitemOpen
  \bibfield  {author} {\bibinfo {author} {\bibfnamefont {C.}~\bibnamefont
  {Alexandrou}}, \bibinfo {author} {\bibfnamefont {S.}~\bibnamefont {Bacchio}},
  \bibinfo {author} {\bibfnamefont {J.}~\bibnamefont {Finkenrath}}, \bibinfo
  {author} {\bibfnamefont {A.}~\bibnamefont {Frommer}}, \bibinfo {author}
  {\bibfnamefont {K.}~\bibnamefont {Kahl}},\ and\ \bibinfo {author}
  {\bibfnamefont {M.}~\bibnamefont {Rottmann}},\ }\bibfield  {title} {\bibinfo
  {title} {{Adaptive Aggregation-based Domain Decomposition Multigrid for
  Twisted Mass Fermions}},\ }\href {https://doi.org/10.1103/PhysRevD.94.114509}
  {\bibfield  {journal} {\bibinfo  {journal} {Phys. Rev. D}\ }\textbf {\bibinfo
  {volume} {94}},\ \bibinfo {pages} {114509} (\bibinfo {year} {2016})},\
  \Eprint {https://arxiv.org/abs/1610.02370} {arXiv:1610.02370 [hep-lat]}
  \BibitemShut {NoStop}%
\bibitem [{\citenamefont {Alexandrou}\ \emph {et~al.}(2019)\citenamefont
  {Alexandrou}, \citenamefont {Bacchio},\ and\ \citenamefont
  {Finkenrath}}]{Alexandrou:2018wiv}%
  \BibitemOpen
  \bibfield  {author} {\bibinfo {author} {\bibfnamefont {C.}~\bibnamefont
  {Alexandrou}}, \bibinfo {author} {\bibfnamefont {S.}~\bibnamefont
  {Bacchio}},\ and\ \bibinfo {author} {\bibfnamefont {J.}~\bibnamefont
  {Finkenrath}},\ }\bibfield  {title} {\bibinfo {title} {{Multigrid approach in
  shifted linear systems for the non-degenerated twisted mass operator}},\
  }\href {https://doi.org/10.1016/j.cpc.2018.10.013} {\bibfield  {journal}
  {\bibinfo  {journal} {Comput. Phys. Commun.}\ }\textbf {\bibinfo {volume}
  {236}},\ \bibinfo {pages} {51} (\bibinfo {year} {2019})},\ \Eprint
  {https://arxiv.org/abs/1805.09584} {arXiv:1805.09584 [hep-lat]} \BibitemShut
  {NoStop}%
\bibitem [{\citenamefont {Clark}\ \emph {et~al.}(2010)\citenamefont {Clark},
  \citenamefont {Babich}, \citenamefont {Barros}, \citenamefont {Brower},\ and\
  \citenamefont {Rebbi}}]{Clark:2009wm}%
  \BibitemOpen
  \bibfield  {author} {\bibinfo {author} {\bibfnamefont {M.~A.}\ \bibnamefont
  {Clark}}, \bibinfo {author} {\bibfnamefont {R.}~\bibnamefont {Babich}},
  \bibinfo {author} {\bibfnamefont {K.}~\bibnamefont {Barros}}, \bibinfo
  {author} {\bibfnamefont {R.~C.}\ \bibnamefont {Brower}},\ and\ \bibinfo
  {author} {\bibfnamefont {C.}~\bibnamefont {Rebbi}},\ }\bibfield  {title}
  {\bibinfo {title} {{Solving Lattice QCD systems of equations using mixed
  precision solvers on GPUs}},\ }\href
  {https://doi.org/10.1016/j.cpc.2010.05.002} {\bibfield  {journal} {\bibinfo
  {journal} {Comput. Phys. Commun.}\ }\textbf {\bibinfo {volume} {181}},\
  \bibinfo {pages} {1517} (\bibinfo {year} {2010})},\ \Eprint
  {https://arxiv.org/abs/0911.3191} {arXiv:0911.3191 [hep-lat]} \BibitemShut
  {NoStop}%
\bibitem [{\citenamefont {Babich}\ \emph {et~al.}(2011)\citenamefont {Babich},
  \citenamefont {Clark}, \citenamefont {Joo}, \citenamefont {Shi},
  \citenamefont {Brower},\ and\ \citenamefont {Gottlieb}}]{Babich:2011np}%
  \BibitemOpen
  \bibfield  {author} {\bibinfo {author} {\bibfnamefont {R.}~\bibnamefont
  {Babich}}, \bibinfo {author} {\bibfnamefont {M.~A.}\ \bibnamefont {Clark}},
  \bibinfo {author} {\bibfnamefont {B.}~\bibnamefont {Joo}}, \bibinfo {author}
  {\bibfnamefont {G.}~\bibnamefont {Shi}}, \bibinfo {author} {\bibfnamefont
  {R.~C.}\ \bibnamefont {Brower}},\ and\ \bibinfo {author} {\bibfnamefont
  {S.}~\bibnamefont {Gottlieb}},\ }\bibfield  {title} {\bibinfo {title}
  {{Scaling Lattice QCD beyond 100 GPUs}},\ }in\ \href
  {https://doi.org/10.1145/2063384.2063478} {\emph {\bibinfo {booktitle} {{SC11
  International Conference for High Performance Computing, Networking, Storage
  and Analysis Seattle, Washington, November 12-18, 2011}}}}\ (\bibinfo {year}
  {2011})\ \Eprint {https://arxiv.org/abs/1109.2935} {arXiv:1109.2935
  [hep-lat]} \BibitemShut {NoStop}%
\bibitem [{\citenamefont {Clark}\ \emph {et~al.}(2016)\citenamefont {Clark},
  \citenamefont {Joó}, \citenamefont {Strelchenko}, \citenamefont {Cheng},
  \citenamefont {Gambhir},\ and\ \citenamefont {Brower}}]{Clark:2016rdz}%
  \BibitemOpen
  \bibfield  {author} {\bibinfo {author} {\bibfnamefont {M.~A.}\ \bibnamefont
  {Clark}}, \bibinfo {author} {\bibfnamefont {B.}~\bibnamefont {Joó}},
  \bibinfo {author} {\bibfnamefont {A.}~\bibnamefont {Strelchenko}}, \bibinfo
  {author} {\bibfnamefont {M.}~\bibnamefont {Cheng}}, \bibinfo {author}
  {\bibfnamefont {A.}~\bibnamefont {Gambhir}},\ and\ \bibinfo {author}
  {\bibfnamefont {R.}~\bibnamefont {Brower}},\ }\bibfield  {title} {\bibinfo
  {title} {{Accelerating Lattice QCD Multigrid on GPUs Using Fine-Grained
  Parallelization}},\ }\href@noop {} {\  (\bibinfo {year} {2016})},\ \Eprint
  {https://arxiv.org/abs/1612.07873} {arXiv:1612.07873 [hep-lat]} \BibitemShut
  {NoStop}%
\bibitem [{\citenamefont {{R Development Core Team}}(2005)}]{R:2005}%
  \BibitemOpen
  \bibfield  {author} {\bibinfo {author} {\bibnamefont {{R Development Core
  Team}}},\ }\href {http://www.R-project.org} {\emph {\bibinfo {title} {{R: A
  language and environment for statistical computing}}}},\ \bibinfo
  {organization} {R Foundation for Statistical Computing},\ \bibinfo {address}
  {Vienna, Austria} (\bibinfo {year} {2005}),\ \bibinfo {note} {{ISBN}
  3-900051-07-0}\BibitemShut {NoStop}%
\bibitem [{\citenamefont {Sheikholeslami}\ and\ \citenamefont
  {Wohlert}(1985)}]{Sheikholeslami:1985ij}%
  \BibitemOpen
  \bibfield  {author} {\bibinfo {author} {\bibfnamefont {B.}~\bibnamefont
  {Sheikholeslami}}\ and\ \bibinfo {author} {\bibfnamefont {R.}~\bibnamefont
  {Wohlert}},\ }\bibfield  {title} {\bibinfo {title} {{Improved Continuum Limit
  Lattice Action for QCD with Wilson Fermions}},\ }\href
  {https://doi.org/10.1016/0550-3213(85)90002-1} {\bibfield  {journal}
  {\bibinfo  {journal} {Nucl. Phys. B}\ }\textbf {\bibinfo {volume} {259}},\
  \bibinfo {pages} {572} (\bibinfo {year} {1985})}\BibitemShut {NoStop}%
\bibitem [{Note3()}]{Note3}%
  \BibitemOpen
  \bibinfo {note} {P. Dimopoulos, private communication.}\BibitemShut {Stop}%
\bibitem [{\citenamefont {Frezzotti}\ and\ \citenamefont
  {Rossi}(2004{\natexlab{c}})}]{Frezzotti:2004wz}%
  \BibitemOpen
  \bibfield  {author} {\bibinfo {author} {\bibfnamefont {R.}~\bibnamefont
  {Frezzotti}}\ and\ \bibinfo {author} {\bibfnamefont {G.~C.}\ \bibnamefont
  {Rossi}},\ }\bibfield  {title} {\bibinfo {title} {{Chirally improving Wilson
  fermions. II. Four-quark operators}},\ }\href
  {https://doi.org/10.1088/1126-6708/2004/10/070} {\bibfield  {journal}
  {\bibinfo  {journal} {J. High Energy Phys.{}}\ }\textbf {\bibinfo {volume}
  {10}},\ \bibinfo {pages} {070} (\bibinfo {year} {2004}{\natexlab{c}})},\
  \Eprint {https://arxiv.org/abs/hep-lat/0407002} {arXiv:hep-lat/0407002
  [hep-lat]} \BibitemShut {NoStop}%
\bibitem [{\citenamefont {Ottnad}\ and\ \citenamefont
  {Urbach}(2018)}]{Ottnad:2017bjt}%
  \BibitemOpen
  \bibfield  {author} {\bibinfo {author} {\bibfnamefont {K.}~\bibnamefont
  {Ottnad}}\ and\ \bibinfo {author} {\bibfnamefont {C.}~\bibnamefont {Urbach}}
  (\bibinfo {collaboration} {ETM}),\ }\bibfield  {title} {\bibinfo {title}
  {{Flavor-singlet meson decay constants from $N_f=2+1+1$ twisted mass lattice
  QCD}},\ }\href {https://doi.org/10.1103/PhysRevD.97.054508} {\bibfield
  {journal} {\bibinfo  {journal} {Phys. Rev. D}\ }\textbf {\bibinfo {volume}
  {97}},\ \bibinfo {pages} {054508} (\bibinfo {year} {2018})},\ \Eprint
  {https://arxiv.org/abs/1710.07986} {arXiv:1710.07986 [hep-lat]} \BibitemShut
  {NoStop}%
\bibitem [{\citenamefont {Hasenfratz}\ \emph {et~al.}(2008)\citenamefont
  {Hasenfratz}, \citenamefont {Hoffmann},\ and\ \citenamefont
  {Schaefer}}]{Hasenfratz:2008fg}%
  \BibitemOpen
  \bibfield  {author} {\bibinfo {author} {\bibfnamefont {A.}~\bibnamefont
  {Hasenfratz}}, \bibinfo {author} {\bibfnamefont {R.}~\bibnamefont
  {Hoffmann}},\ and\ \bibinfo {author} {\bibfnamefont {S.}~\bibnamefont
  {Schaefer}},\ }\bibfield  {title} {\bibinfo {title} {{Reweighting towards the
  chiral limit}},\ }\href {https://doi.org/10.1103/PhysRevD.78.014515}
  {\bibfield  {journal} {\bibinfo  {journal} {Phys. Rev. D}\ }\textbf {\bibinfo
  {volume} {78}},\ \bibinfo {pages} {014515} (\bibinfo {year} {2008})},\
  \Eprint {https://arxiv.org/abs/0805.2369} {arXiv:0805.2369 [hep-lat]}
  \BibitemShut {NoStop}%
\bibitem [{\citenamefont {Finkenrath}\ \emph {et~al.}(2013)\citenamefont
  {Finkenrath}, \citenamefont {Knechtli},\ and\ \citenamefont
  {Leder}}]{Finkenrath:2013soa}%
  \BibitemOpen
  \bibfield  {author} {\bibinfo {author} {\bibfnamefont {J.}~\bibnamefont
  {Finkenrath}}, \bibinfo {author} {\bibfnamefont {F.}~\bibnamefont
  {Knechtli}},\ and\ \bibinfo {author} {\bibfnamefont {B.}~\bibnamefont
  {Leder}},\ }\bibfield  {title} {\bibinfo {title} {{One flavor mass
  reweighting in lattice QCD}},\ }\href
  {https://doi.org/10.1016/j.nuclphysb.2013.10.019,
  10.1016/j.nuclphysb.2014.01.019} {\bibfield  {journal} {\bibinfo  {journal}
  {Nucl. Phys. B}\ }\textbf {\bibinfo {volume} {877}},\ \bibinfo {pages} {441}
  (\bibinfo {year} {2013})},\ \bibinfo {note} {[Erratum: Nucl. Phys. B
  \textbf{880}, 574 (2014)]},\ \Eprint {https://arxiv.org/abs/1306.3962}
  {arXiv:1306.3962 [hep-lat]} \BibitemShut {NoStop}%
\bibitem [{\citenamefont {Leder}\ and\ \citenamefont
  {Finkenrath}(2015)}]{Leder:2015fea}%
  \BibitemOpen
  \bibfield  {author} {\bibinfo {author} {\bibfnamefont {B.}~\bibnamefont
  {Leder}}\ and\ \bibinfo {author} {\bibfnamefont {J.}~\bibnamefont
  {Finkenrath}},\ }\bibfield  {title} {\bibinfo {title} {{Tuning of the strange
  quark mass with optimal reweighting}},\ }\bibfield  {booktitle} {\emph
  {\bibinfo {booktitle} {{Proceedings, 32nd International Symposium on Lattice
  Field Theory (Lattice 2014): Brookhaven, NY, USA, June 23-28, 2014}}},\
  }\href {https://doi.org/10.22323/1.214.0040} {\bibfield  {journal} {\bibinfo
  {journal} {PoS}\ }\textbf {\bibinfo {volume} {LATTICE2014}},\ \bibinfo
  {pages} {040} (\bibinfo {year} {2015})},\ \Eprint
  {https://arxiv.org/abs/1501.06617} {arXiv:1501.06617 [hep-lat]} \BibitemShut
  {NoStop}%
\bibitem [{\citenamefont {Finkenrath}\ \emph {et~al.}(2015)\citenamefont
  {Finkenrath}, \citenamefont {Knechtli},\ and\ \citenamefont
  {Leder}}]{Finkenrath:2015ava}%
  \BibitemOpen
  \bibfield  {author} {\bibinfo {author} {\bibfnamefont {J.}~\bibnamefont
  {Finkenrath}}, \bibinfo {author} {\bibfnamefont {F.}~\bibnamefont
  {Knechtli}},\ and\ \bibinfo {author} {\bibfnamefont {B.}~\bibnamefont
  {Leder}},\ }\bibfield  {title} {\bibinfo {title} {{Isospin Effects by Mass
  Reweighting}},\ }\bibfield  {booktitle} {\emph {\bibinfo {booktitle}
  {{Proceedings, 32nd International Symposium on Lattice Field Theory (Lattice
  2014): Brookhaven, NY, USA, June 23-28, 2014}}},\ }\href
  {https://doi.org/10.22323/1.214.0297} {\bibfield  {journal} {\bibinfo
  {journal} {PoS}\ }\textbf {\bibinfo {volume} {LATTICE2014}},\ \bibinfo
  {pages} {297} (\bibinfo {year} {2015})},\ \Eprint
  {https://arxiv.org/abs/1501.06441} {arXiv:1501.06441 [hep-lat]} \BibitemShut
  {NoStop}%
\end{thebibliography}%

\end{document}